\documentclass[prc,twoside,superscriptaddress,showpacs]{revtex4}
\usepackage{graphicx,epsfig,latexsym,amssymb}
\usepackage{multirow,amsmath,array,booktabs}
\usepackage[section]{placeins}

\newcommand{\ben}{\begin{enumerate}}
\newcommand{\een}{\end{enumerate}}

\newcommand{\beq}{\begin{equation}}
\newcommand{\eeq}{\end{equation}}
\newcommand{\beqn}{\begin{eqnarray}}
\newcommand{\eeqn}{\end{eqnarray}}

\newcommand{\beqd}{\begin{eqnarray*}}
\newcommand{\eeqd}{\end{eqnarray*}}
\newcommand{\bea}{\begin{array}}
\newcommand{\eea}{\end{array}}
\newcommand{\bcen}{\begin{center}}
\newcommand{\ecen}{\end{center}}
\newcommand{\btab}{\begin{tabular}}
\newcommand{\etab}{\end{tabular}}
\newcommand{\bsub}{\begin{subequations}}
\newcommand{\esub}{\end{subequations}}
\newcommand{\bit}{\begin{itemize}}
\newcommand{\eit}{\end{itemize}}
\newcommand{\brule}{\begin{ruledtabular}}
\newcommand{\erule}{\end{ruledtabular}}
\newcommand{\bpm}{\begin{pmatrix}}
\newcommand{\epm}{\end{pmatrix}}

\newcommand{\cals}[1]{{\cal #1}}

\newcommand{\ba}{\mbox{\boldmath$a$}}

\newcommand{\ff}[1]{\frac{1}{#1}}
\newcommand{\lc}{\left<}
\newcommand{\rc}{\right>}
\newcommand{\lr}{\left|}
\newcommand{\rl}{\right|}
\newcommand{\lb}{\left(}
\newcommand{\rb}{\right)}
\newcommand{\ls}{\left[}
\newcommand{\rs}{\right]}
\newcommand{\Lb}{\left\{}
\newcommand{\Rb}{\right\}}

\renewcommand{\emph}[1]{\textit{\textbf{#1}}}
\newcommand{\svec}[1]{{\mbox{\boldmath${\rm #1}$}}}
\newcommand{\ivec}{\vec}

\newcommand{\fm}{\text{fm}}
\newcommand{\mev}{\text{MeV}}
\newcommand{\cm}{{\text{c.m.}}}

\begin{document}
\title{New effective interactions in RMF theory with non-linear terms
and density-dependent meson-nucleon coupling
  \footnote
  {Supported partly by the Major State Basic Research Development Program under
  Contract Number G2000077407 and the National Natural Science Foundation
  of China under Grant Nos. 19847002 and 19935030.} }
\author{Wenhui Long}
\affiliation{School of Physics, Peking University, 100871 Beijing,
China} \affiliation{Institut de Physique Nucl$\acute{e}$aire,
CNRS-IN2P3, Universit$\acute{e}$ Paris-Sud, 91406 Orsay, France}
\author{Jie Meng }
\email[]{mengj@pku.edu.cn}
\affiliation{School of Physics, Peking University, 100871 Beijing, China}
\affiliation{Institute of Theoretical Physics, Chinese Academy of Sciences, Beijing, China}
\affiliation{Center of Theoretical Nuclear Physics, National Laboratory of
       Heavy Ion Accelerator, 730000 Lanzhou, China}
\author{Nguyen Van Giai}
\affiliation{Institut de Physique Nucl$\acute{e}$aire, CNRS-IN2P3,
Universit$\acute{e}$ Paris-Sud, 91406 Orsay, France}
\author{Shan-Gui Zhou}
\affiliation{School of Physics, Peking University, 100871 Beijing, China}
\affiliation{Max-Planck-Institut f\"ur Kernphysik, 69029 Heidelberg, Germany}
\affiliation{Institute of Theoretical Physics, Chinese Academy of Sciences, Beijing, China}
\affiliation{Center of Theoretical Nuclear Physics, National Laboratory of
       Heavy Ion Accelerator, 730000 Lanzhou, China}

\begin{abstract}\vspace{2em}

New parameter sets for the Lagrangian density in the relativistic
mean field (RMF) theory, PK1 with nonlinear $\sigma$- and
$\omega$-meson self-coupling, PK1R with nonlinear $\sigma$-,
$\omega$- and $\rho$-meson self-coupling and PKDD with the
density-dependent meson-nucleon coupling, are proposed. They are
able to provide an excellent description not only for the
properties of  nuclear matter but also for the nuclei in and far
from the valley of beta-stability. For the first time in the
parametrization of the RMF Lagrangian density, the center-of-mass
correction is treated by a microscopic way, which is essential to
unify the description of nuclei from light to heavy regions with
one effective interaction.

\end{abstract}


\maketitle

\section{Introduction}
In the past decade, the development of unstable nuclear
beams~\cite{Tan:1995,Mul:2001} has extended our knowledge of
nuclear physics from the stable nuclei and those nearby to the
unstable nuclei far from the stability line. Intense research in
this area shows that there exist lots of unexpected phenomena:
strange nuclear structure like neutron halo (skin) and proton halo
(skin)~\cite{Tan:1985,Tan:1988,Warn:1995,Chul:1996,Meng:1996,Meng:1998,Meng:1998PLB,Meng:2002PLB},
soft excitation modes~\cite{Koba:1989, Koba:1992}, the enhancement
of fusion cross sections induced by the extended matter
distributions~\cite{Alm:1995,Yosh:1995} etc.. With further
developments, many other new features will be found. It also
becomes very important to find a reliable theory and improve the
reliability for predicting the properties of even more exotic
nuclei out to the proton and neutron drip lines.

Relativistic mean field (RMF)~\cite{Wale:1974,Serot:1986} theory
has received wide attention because of its successful description
of many nuclear phenomena during the past years. With a very
limited number of parameters, RMF theory is able to give a
satisfactory description for the ground state properties of
spherical~\cite{Reinhard:1989} and deformed nuclei
~\cite{Ring:1996} at and away from the stability line.  The recent
reviews on RMF theory can be seen in
~\cite{Serot:1986,Reinhard:1989,Ring:1996}. In the simplest
version of RMF theory, the mesons do not interact among
themselves, which results in a too large incompressibility for
nuclear matter. Boguta and Bodmer~\cite{Boguta:1977} therefore
proposed to include a nonlinear self-coupling of the
$\sigma$-field, a concept which has been used in almost all the
recent applications. The meson self-coupling introduces a new
density dependence into the Lagrangian and consequently, the
nuclear matter incompressibility can be lowered to reasonable
values. As an implement, in 1994 the nonlinear self-coupling of
the $\omega$-field is introduced by Sugahara and
Toki~\cite{Sugahara:1994}. In this paper we will introduce the
nonlinear self-coupling for the $\rho$-field. Recently RMF theory
with density-dependent (DD) meson-nucleon
couplings~\cite{Lenske:1995,Fuchs:1995,Typel:1999,Niksic:2002} was
developed by various authors.

Till now the two versions (the nonlinear self-coupling of meson
fields and the DD meson-nucleon couplings) of RMF theory have been
successfully applied to describe the nuclear properties, including
binding energies, nuclear matter distribution, single-particle
spectra, magnetic moments, collective excited states, dipole sum
rule, shell effects in finite nuclei, pseudospin symmetry,
rotating nuclei, superdeformed bands, etc. In particular, the halo
phenomena can be understood self-consistently in this
microscopical model after the proper treatment of the continuum
effect~\cite{Meng:1996,Meng:1998a}. Combining with Glauber model,
the charge-changing cross sections for C, N, O and F isotopes are
calculated and good agreement with the data has been
achieved\cite{Meng:1998PLB,Meng:2002PLB}. The long existing
problem for the origin of pseudo-spin symmetry in nuclei is given
naturally as a relativistic symmetry
\cite{Ginocchio:1997,Meng:1998PRCps,Meng:1999PRC}. Good agreement
with experimental data has also been found recently for magnetic
rotation \cite{Madokoro:2002PRC}, collective excitations such as
giant resonances~\cite{Ma:2002} and for twin bands in rotating
super-deformed nuclei ~\cite{Konig:1993}. It is also noted that
cranked RMF theory provides an excellent description of
superdeformed rotational bands in the A=140-150
region~\cite{Afanasjev:1996a}, in the Sr
region~\cite{Afanasjev:1996} and in the Hg region
~\cite{Konig:1996}.

Among the existing parametrizations for RMF theory, the most
frequently used are NL1~\cite{Reinhard:1986}, PL-40
~\cite{Reinhard:1988}, NL-SH~\cite{Sharma:1993}, TM1
~\cite{Sugahara:1994} and NL3~\cite{Lalazissis:1997} with
nonlinear self-coupling of mesons, and TW99~\cite{Typel:1999} and
DD-ME1~\cite{Niksic:2002} with DD meson-nucleon coupling. The
effective interactions NL1, NL3, TM1,NL-SH and TW99, DD-ME1 give
good results in most of the cases.

Along the beta stability line NL1 gives excellent results for
binding energies and charge radii, in addition it provides an
excellent description of the superdeformed bands
~\cite{Afanasjev:1996a,Afanasjev:1996}. However, in going away
from the stability line the results are less satisfactory. This
can be partly attributed to the large asymmetry energy $J\simeq$
44 MeV predicted by this force. In addition, the calculated
neutron skin thickness shows systematic deviations from the
experimental values for the set NL1.

In the effective interaction NL-SH this problem was treated in a
better way and the improved isovector properties have been
obtained with an asymmetry energy of $J\simeq$ 36 MeV. Moreover,
NL-SH seems to describe the deformation properties in a better way
than NL1. However, the NL-SH parametrization produces a slight
over-binding along the line of beta-stability and in addition it
fails to reproduce successfully the superdeformed minima in
Hg-isotopes in constrained calculations for the energy landscape.
A remarkable disagreement between the two parametrizations are the
quite different values predicted for the nuclear matter
incompressibility. NL1 predicts a small value ($K$ = 212 MeV)
while with NL-SH a very large value ($K$ = 355 MeV) is obtained.
Both forces fail to reproduce the experimental values for the
isoscalar giant monopole resonances for Pb and Zr nuclei. The NL1
parametrization underestimates the empirical data by about 2 MeV
while NL-SH overestimates it by about the similar values. As an
improvement, the effective interactions, NL3 and TM1, provide
reasonable compression modulus ($K_{\text{NL3}}=268.0\ \mev,
K_{\text{TM1}}=281.16\ \mev$) and asymmetry energy
($J_{\text{NL3}}=36.56\ \mev, J_{\text{TM1}}=36.89\ \mev$) but
fairly small baryonic saturation density
($\rho_{\text{NL3}}=0.145, \rho_{\text{TM1}}=0.145$). In order to
improve the description of these quantities, we developed two
nonlinear self-coupling parametrizations called as PK1 with
nonlinear $\sigma$- and $\omega$-meson nonlinear self-coupling,
and PK1R with nonlinear $\sigma$-, $\omega$- and $\rho$-meson
nonlinear self-coupling.

RMF theory with DD meson-nucleon couplings
~\cite{Lenske:1995,Fuchs:1995} is an alternative approach to the
description of nuclear matter and finite nuclei as compared to the
model with the nonlinear self-interactions of mesons. There exist
two representative parametrizations, TW99~\cite{Typel:1999} and
DD-ME1~\cite{Niksic:2002} for the density-dependence of
meson-nucleon coupling. They are able to describe quantitatively
properties of nuclear matter and finite nuclei with similar
quality as the parametrizations of the nonlinear self-coupling. As
a comparison with the nonlinear self-coupling, we also developed a
parametrization named as PKDD with DD meson-nucleon couplings.

\begin{figure}[htbp]
\includegraphics[width = 12.0cm]{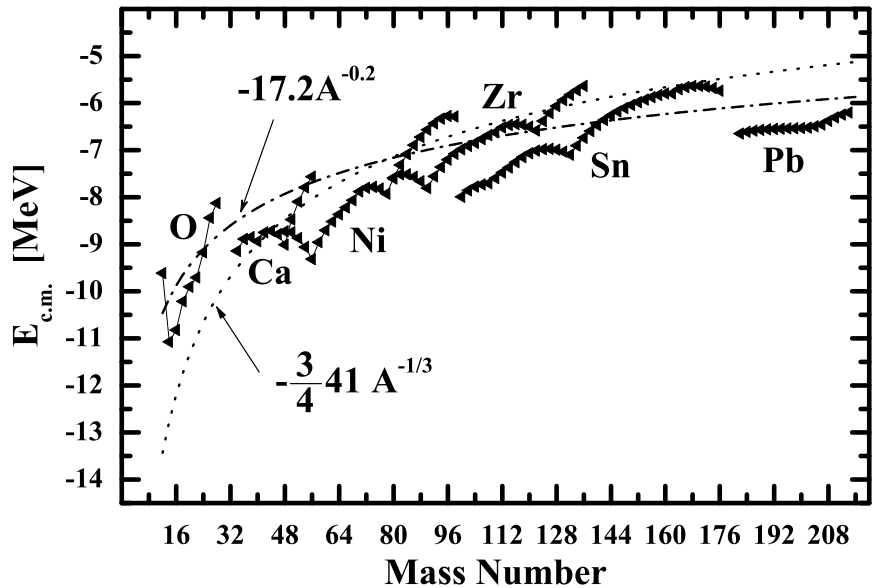}
\caption{The microscopic  center-of-mass correction, in comparison
with two phenomenological cases.
 }\label{fig:ecm-all2}
\end{figure}

In all previous nonlinear self-coupling parametrizations as
mentioned above, the center-of-mass correction is made by a
phenomenological way. In the present parametrizations, the
PK-series, the contribution from the center-of-mass motion is
treated in a microscopic way~\cite{Bender:2000}. The systematic
behavior of the center-of-mass correction on nuclear masses is
shown in Fig.~\ref{fig:ecm-all2}. From this graph, we can see that
it is essential to choose a proper method to treat the
center-of-mass motion for both light and heavy nuclei. Obviously,
the microscopic method provides more reasonable and reliable
results for the center-of-mass motion (for the details, see
Sec.~\ref{sec:3}).

In section~\ref{sec:2}, we will present a short summary of RMF
theory with the nonlinear self-coupling of meson fields and with
DD meson-nucleon coupling and the relationship between them. The
details of our parametrizations are given in section~\ref{sec:3}.
In section~\ref{sec:4} and~\ref{sec:5}, we study the bulk
properties of nuclear matter and spherical nuclei with the newly
obtained effective interactions. The detailed microscopic
structure of doubly magic nuclei is also investigated in the end
of section~\ref{sec:5}. Finally conclusions are given in section
\ref{sec:6}.

\section{RMF Theory with Nonlinear Self-coupling and Density-dependent Meson-nucleon Coupling}
\label{sec:2}

The basic ansatz of the RMF theory~\cite{Serot:1986} is a
Lagrangian density whereby nucleons are described as Dirac
particles which interact via the exchange of various mesons and
the photon. The mesons considered are the scalar sigma ($\sigma$),
vector omega ($\bf \omega$) and iso-vector vector rho ($\ivec
\rho$).  The latter provides the necessary isospin asymmetry.  The
Lagrangian then consists of the free baryon and meson parts and
the interaction part with minimal coupling, together with the
nucleon mass $M$ and $m_\sigma$ ($g_\sigma$), $m_\omega$
($g_\omega$) and $m_\rho$ ($g_\rho$) the masses (coupling
constants) of the respective mesons: \beq\label{lagrangian}
\begin{split}
\cals L=&\bar{\psi}\ls i{\gamma^\mu}{\partial_\mu}-M-{g_\sigma}\sigma
 - g_\omega\gamma^\mu\omega_\mu - g_\rho \gamma^\mu {\ivec
\tau}\cdot \ivec\rho_\mu - e\gamma^\mu\frac {1-\tau_3}{2}A_\mu \rs\psi\\
&+\frac{1}{2}\partial^\mu\sigma\partial_\mu\sigma-\frac{1}{2}m_\sigma^2\sigma^2
-\frac{1}{3}g_2\sigma^3-\frac{1}{4}g_3\sigma^4
-\frac{1}{4}\omega^{\mu\nu}\omega_{\mu\nu}
+\frac{1}{2}m_\omega^2\omega^\mu\omega_\mu+\frac{1}{4}c_3\lb\omega^\mu\omega_\mu\rb^2\\
&-\frac{1}{4}\ivec\rho^{\mu\nu}\cdot\ivec\rho_{\mu\nu}
+\frac{1}{2}m_\rho^2\ivec\rho^\mu\cdot\ivec\rho_\mu
+\frac{1}{4}d_3\lb\ivec\rho^\mu\cdot\ivec\rho_\mu\rb^2
-\frac{1}{4}A^{\mu\nu}A_{\mu\nu},
\end{split}
\eeq
where the tensor quantities are
\bsub\beqn
\omega^{\mu\nu}&=&\partial^\mu\omega^\nu-\partial^\nu\omega^\mu\\
\ivec\rho^{\mu\nu}&=&\partial^\mu\ivec\rho^\nu-\partial^\nu\ivec\rho^\mu
+ g_\rho\ivec\rho^\mu\times\ivec\rho^\nu\\
A^{\mu\nu}&=&\partial^\mu A^\nu -\partial^\nu A^\mu.
\eeqn\esub

In this paper we use arrow for isospin vectors and bold type for
space vectors, respectively. There are 11 parameters in the
Lagrangian density (\ref{lagrangian}), i.e., 4 masses ($M,
m_\sigma, m_\omega, m_\rho$), 3 nucleon-meson coupling constants
($g_\sigma, g_\omega, g_\rho$), and 4 self-coupling constants
($g_2, g_3, c_3,d_3$). Generally, the nucleon mass $M$ and
$\rho$-meson mass $m_\rho$ ( also sometimes the mass of
$\omega$-meson in some parameterizations ) are fixed to their free
values and the nonlinear coupling coefficients $c_3$ and $d_3$ are
taken as zeros. Then the remaining 6$\sim$7 parameters are
determined by the fitting to the experimental observables.

Different from the nonlinear self-coupling version, the
meson-nucleon interactions are described as density-dependent in
the DD meson-nucleon coupling RMF theory~\cite{Typel:1999,
Niksic:2002}. The meson-nucleon coupling constants $g_\sigma,
g_\omega, g_\rho$ become functions of the baryonic density
$\rho_v$, $\rho_v=\sqrt{j_\mu j^\mu} $ where $j^\mu
=\bar\psi\gamma^\mu\psi$ and the nonlinear self-coupling constants
$g_2,g_3,c_3,d_3$ are set to zero in the Lagrangian density
(\ref{lagrangian}).

For $\sigma$ and $\omega$-meson, the baryonic density-dependence
of the coupling constants is adopted as
 \beq
 g_i(\rho_v) = g_i(\rho_{\text{sat.}}) f_i(x)~~~~~~~~\text{ for } i =\sigma, \omega
 \eeq
where
 \beq
 f_i(x) = a_i\frac{1 + b_i(x+d_i)^2}{1+ c_i(x+d_i)^2}
 \eeq
is a function of $x =\rho_v/\rho_{\text{sat.}}$, and
$\rho_{\text{sat.}}$ denotes the baryonic saturation density of
nuclear matter.

For the $\rho$ meson, an exponential dependence is utilized as
\beq g_\rho  = g_\rho(\rho_{\text{sat.}}) \exp[-a_\rho(x-1)] \eeq

For the functions $f_i(x)$, one has five constraint conditions
$f_i(1) = 1, f_\sigma''(1) = f_\omega''(1)$ and $f_i''(0) =0$.
Then 8  parameters related to density dependence for $\sigma$-N
and $\omega$-N couplings are reduced to 3 free parameters. As
mentioned above, the masses of nucleon and $\rho$-meson are fixed
in general and the nonlinear self-coupling constants $g_2, g_3,
c_3$ and $d_3$ are set to zero. With 4 free parameters for density
dependence, there totally are 8$\sim$9 parameters left free in the
Lagrangian density (\ref{lagrangian}) for the density-dependent
meson-nucleon coupling RMF theory.

The single-nucleon Dirac equation is derived by the variation of
the Lagrangian density (\ref{lagrangian}) with respect to
$\bar\psi$,
 \beq\label{Dirac}
 \ls i\gamma^\mu\partial_\mu - (M+\Sigma_S) -\gamma^\mu\Sigma_\mu\rs\psi=0
 \eeq
with the nucleon self-energies $\Sigma_\mu$ and $\Sigma_S$ defined
by the following relations:
 \bsub\label{potentials}
 \beqn
\Sigma_S   &=& g_\sigma\sigma\\
\Sigma_\mu &=&
g_\omega\omega_\mu + g_\rho\ivec\tau\cdot\ivec\rho_\mu + e\frac{1-\tau_3}{2}A_\mu
+\Sigma_\mu^R
 \eeqn
 \esub
where the rearrangement term $\Sigma_\mu^R$ comes from the
density-dependence of the meson-nucleon coupling constants,
 \beq\label{rearrange}
\Sigma_R^\mu=\frac{j^\mu}{\rho_v}\lb\frac{\partial g_\omega}{\partial \rho_v}
\bar\psi\gamma^\nu\psi\omega_\nu+\frac{\partial g_\rho}{\partial \rho_v}
\bar\psi\gamma^\nu\ivec\tau\psi\cdot\ivec\rho_\nu+\frac{\partial
g_\sigma}{\partial\rho_v} \bar\psi\psi\sigma\rb
 \eeq
which is reduced to zero in  RMF theory with the nonlinear self-coupling.

The Klein-Gordon equations for mesons are obtained by the
variation of the Lagrangian density (\ref{lagrangian}) with
respect to the corresponding meson field operators
 \bsub\label{Klein-Gordon}
\beqn
\ls-\Delta + m_\sigma\rs \sigma &=& - g_\sigma\rho_s - g_2\sigma^2 - g_3\sigma^3\\
\ls-\Delta + m_\omega\rs \omega &=&   g_\omega\rho_b - c_3\omega^3\\
\ls-\Delta + m_\rho  \rs \rho   &=&   g_\rho\ls\rho_b^{(n)}-\rho_b^{(p)}\rs - d_3\rho^3
\eeqn
 \esub
For the RMF theory with the nonlinear self-coupling, the self-coupling
of mesons can be expressed into the density-dependence of  meson-nucleon coupling by
redefining the coupling constants,
 \bsub\label{couplings}
\beqn
g_\sigma &=&  g_\sigma + \ls g_2\sigma^2 + g_3\sigma^3\rs/\rho_s\\
g_\omega &=& g_\omega - c_3\omega^3/\rho_b\\
g_\rho   &=& g_\rho - d_3\rho^3/\ls\rho_b^n-\rho_b^p \rs
\eeqn
 \esub
The behaviors of the coupling constants with respect to the
baryonic density are shown in Fig.~\ref{fig:gsig&gome}.

In order to compare with the experimental mass of a nucleus, the
calculated nuclear mass
 \beq
M = M_{\text{rmf}} + E_\cm + E_{\text{pair}} \eeq is obtained from
the mean field contribution by adding corrections due to the
center-of-mass motion and pairing effects for open shell nuclei.
The mean field contribution $M_{\text{rmf}}$ is derived from a
spatial integration of the "00" component of the energy-momentum
tensor. In the nonlinear (N.L.) self-coupling case, it reads
 \beq\label{ene1}
\begin{split}
M_{\text{rmf}}^{\text{N.L.}}=&\sum_{a=1}^A\varepsilon_a\lc\bar\phi_a\gamma^0\phi_a
\rc
+\int_V d^3r\Lb-\ff2g_\sigma\sigma\rho_s-\ff6 g_2\sigma^3-\ff4g_3\sigma^4\Rb\\
&+\int_V d^3r\Lb-\ff2 g_\omega\omega\rho_b
+\frac{1}{4}c_3\omega^4-\ff2g_\rho\rho\rho_b^{(3)}+\frac{1}{4}d_3\rho^4-\ff2
eA_0\rho_b^p\Rb
\end{split}
 \eeq
where $\phi_a$ denotes the single-particle spinor in the nucleus.
In the density-dependent (D.D.) case, the high-order($\ge 3$)
terms of mesons in (\ref{ene1}) are replaced by a "rearrangement"
term,
 \beq\label{ene2}
\begin{split}
M_{\text{rmf}}^{\text{D.D.}}=&\sum_{a=1}^A\varepsilon_a\lc\bar\phi_a\gamma^0\phi_a \rc
+\int_V d^3r\Lb-\ff2g_\sigma\sigma\rho_s-\ff2 g_\omega\omega\rho_b-\ff2g_\rho\rho\rho_b^{(3)}
-\ff2 eA_0\rho_b^p - \Sigma_R\rho_b\Rb.
\end{split}
 \eeq
where
 \beq
\Sigma_R=\ff{\rho_{\text{sat.}}}\ls
g_\omega(\rho_{\text{sat.}})f'_\omega(x)\rho_0 \omega - a_\rho
g_\rho(\rho_{\text{sat.}})e^{ -a_\rho(x-1)} \rho_0^{(3)}\rho_3 +
g_\sigma(\rho_{\text{sat.}})f'_\sigma(x)\rho_s \sigma  \rs
 \eeq
is the time-component of $\Sigma^\mu_R$ (see (\ref{rearrange})).

The correction from the center-of-mass motion is calculated from
the projection-after-variation in first-order
approximation~\cite{Bender:2000}:
 \beq \label{ecm1}
E^{\text{mic.}}_\cm = -\frac{1}{2MA}\lc {\svec P}_\cm^2\rc
 \eeq
where the center-of-mass momentum $\svec P_\cm =\sum_i^A\svec p_i$
and the expectation value of its square $\lc \hat{\svec
P}_\cm^2\rc$  reads
 \beq\label{ecm2}
 \lc {\svec P}_\cm^2\rc = \sum_a v_a^2 p_{aa}^2 -\sum_{a,b} v_a^2 v_b^2\svec
p_{ab}\cdot\svec p^*_{ab} +\sum_{a,b}v_a u_a v_b u_b \svec
p_{ab}\cdot\svec p_{\bar a\bar b}
 \eeq
with occupation probabilities $v_a^2$ and $u_a^2 = 1-v_a^2$
accounting for pairing effects, where $a,b$ denote the BCS states
(see below).

The contribution from the pairing correlations  $E_{\text{pair}}$
are treated in the BCS approximation
 \beq
E_{\text{pair}} = -\Delta\sum_a v_a u_a
 \eeq
with the pairing gap $\Delta$ taken from the calculation of
relativistic continuum Hartree-Bogoliubov (RCHB) theory with
zero-range pairing interaction~\cite{Meng:1998a}.

For the nuclear radii, the effects from the center-of-mass motion
are also taken into account as follows. Because of its fairly
small effects, a rather rough correction is adopted for protons
 \beq\label{radii-cm}
\delta R_p^2 = -\frac{2}{Z}\sum_a^A \lc\phi_a\rl \svec R_\cm\cdot\sum_{i}^Z \svec r_i
\lr \phi_a\rc +\sum_a^A \lc\phi_a\rl\svec R_\cm^2\lr \phi_a\rc
 \eeq
where the center-of-mass coordinate $\svec R_\cm =1/A \sum_i ^A \svec r_i$. Then we get
\beq\label{cmc-r}
\delta R_p^2 = -\frac{2}{A} R_p^2 + \ff A R_M^2
 \eeq
where $R_p $ and $R_M$ denote the proton and matter radii. Here we
only consider the direct-term contributions in (\ref{radii-cm}) to
keep with the spirit of RMF theory. For the neutron radii, we use
the same procedure as for protons. The charge radius is obtained
from the proton radius combining with the proton and neutron size
and the center-of-mass correction (\ref{cmc-r}) is included in
$R_p^2$~\cite{Sugahara:1994}
 \beq\label{rch}
R_{\text{ch}}^2 = R_p^2 + (0.862\ \fm)^2 - (0.336\ \fm)^2 N/Z
 \eeq

\section{Parametrization of Effective Lagrangian and Numerical Details}
\label{sec:3}

The aim of the present investigation is to provide new improved
effective interactions for the Lagrangian density
(\ref{lagrangian}) with the nonlinear self-coupling and
density-dependent meson-nucleon coupling in RMF theory. A
multi-parameter fitting was performed with the Levenberg-Marquardt
method~\cite{Press:1992}. Two nonlinear self-coupling effective
interactions  have been obtained, PK1 with $\sigma$- and
$\omega$-meson self-coupling, and PK1R with $\sigma$-, $\omega$-
and $\rho$-meson self-coupling (see Table~\ref{tab:psets}). A
density-dependent meson-nucleon coupling effective interaction has
also been obtained with PKDD (see Table~\ref{tab:psets} and
\ref{tab:pkdd}). In our parametrization, the masses of neutron and
proton are fixed to their free values: $M_n=939.5731\ \mev,
M_p=938.2796\ \mev$, and the mass of $\rho$-meson is fixed to its
experimental value $763.0\ \mev$. The mass of $\omega$-meson is
slightly adjusted in obtaining the effective interactions PK1 and
PK1R while fixed to $783.0\ \mev$ for the effective interaction
PKDD.

The masses of the spherical nuclei $^{16}$O, $^{40}$Ca, $^{48}$Ca,
$^{56}$Ni, $^{68}$Ni, $^{90}$Zr, $^{116}$Sn, $^{132}$Sn,
$^{194}$Pb and $^{208}$Pb are fitted to give the effective
interactions PK1, PK1R and PKDD. The experimental inputs for
finite nuclei used in the fitting procedure are shown in Table
\ref{tab:ebn-all}. These inputs have been used to minimize the
least square error:
 \beq \label{chisquare} \chi^2(\ba) =
\sum_{i=1}^N \ls \frac{y^{exp}_i-y(x_i;\ba)} {\sigma_i}\rs^2
 \eeq
where $\ba$ is the ensemble of parameters to be fitted,
$y^{exp}_i$ and $\sigma_i$ are the experimental observable and
corresponding weight.

In the fitting procedure, only the masses of the spherical nuclei
mentioned above and the compression modulus $K$, the baryonic
density at saturation $\rho_{\text{sat.}}$  and the asymmetry
energy $J$ of nuclear matter are included. Here, we should mention
that the radii are excluded because we found that the values of
the compression modulus and  baryonic saturation density are
essential to give a good description of the masses and radii. For
a fixed value of compression modulus, a large baryonic saturation
density will give a small charge radius. By this way, we can
choose the appropriate value for these two quantities to give a
proper description for both mass and charge radius of the chosen
nuclei. To give a fairly precise description on the masses, the
center-of-mass correction is essential for both light and heavy
nuclei. As it can be seen in Fig.~\ref{fig:ecm-all2}, the
deviation between the microscopic and phenomenological results is
considerably large not only for the light nuclei but also for the
heavy ones. We also find that there exist very remarkable shell
effects in the microscopic results which are impossible to obtain
with the phenomenological methods. Hence, we choose the
microscopic center-of-mass correction~\cite{Bender:2000} to deal
with the center-of-mass motion.

The numerical procedure in obtaining the parameter set PK1 is as
follows.
 \ben

\item First, we start from an initial effective interaction and
fix the mass of $\omega$-meson, let the other parameters to be
adjusted by the Levenberg-Marquardt method~\cite{Press:1992}.

\item Basing on the binding energies and charge radii of the
selected nuclei obtained in the first step, we do some adjustment
on the expectation of compression modulus and  baryonic saturation
density to improve the description on the spherical nuclei. Also,
the weights of the observables would be slightly adjusted to
improve the parametrization.

\item  Choose another initial effective interaction and take the
same procedure as in step 2. Then, taking the average of these two
obtained interactions as the new initial effective interaction, we
do the minimizing procedure under the new weights which come from
the previous results.

\item Introduce the adjustment on the mass of $\omega$-meson to
obtain the effective interaction PK1.

\een

Because the contribution  to the nuclear mass from the nonlinear
$\rho$-meson term is fairly small, we fix the nonlinear
self-coupling constant $d_3$ to 350.0 and adjust other parameters
to obtain the effective interaction PK1R.

To obtain the density-dependent meson-nucleon coupling effective
interaction PKDD, we leave the density-dependent parameters
$a_\sigma, d_\sigma$ and $d_\omega$ to be adjusted and the five
others to be determined by the constrain conditions on the
functions $f_i(x), i=\sigma,\omega$. The mass of $\omega$-meson is
fixed to 783.0\mev.

Thus obtained parameter sets are shown in Table~\ref{tab:psets}
and \ref{tab:pkdd} in comparison with other effective interactions
TM1~\cite{Sugahara:1994}, NL3~\cite{Lalazissis:1997},
TW99~\cite{Typel:1999} and DD-ME1~\cite{Niksic:2002}. In
Table~\ref{tab:ebn-all} and~\ref{tab:charge-all} we respectively
list the masses and charge radii of finite nuclei which have been
used in obtaining these seven effective interactions. Compared
with other effective interactions, our newly obtained ones
reproduce well the experimental masses~\cite{Audi:1995}. For these
new effective interactions, only 4$\sim$5 nuclear massed deviate
by more than $1\mev$ (see Table~\ref{tab:ebn-all}). The new
effective interactions PK1, PK1R and PKDD also well describe the
charge radii for these nuclei, especially for Pb isotopes. One can
get a clear idea about the improvement of the new parameter sets
on the description of bulk properties for finite nuclei from the
root of relative square (rrs) deviation $\delta$. In the last row
of Table~\ref{tab:ebn-all} (\ref{tab:charge-all}), the rrs
deviation of the calculated total binding energy (charge radius)
from the data is given. For the total binding energy, the rrs
deviations from the new parameter sets are much smaller than those
from old ones. For the charge radius, the rrs deviations from the
new interactions are comparable with those from NL3 and DD-ME1,
but a bit smaller than those from TM1 and TW99.

Table~\ref{tab:nm-all} lists the nuclear matter quantities
calculated with the newly obtained effective interactions PK1,
PK1R and PKDD, in comparison with other interactions. All the new
effective interactions give proper value of the compression
modulus $K$. Compared with the nonlinear self-coupling effective
interactions TM1~\cite{Sugahara:1994} and
NL3~\cite{Lalazissis:1997}, the new ones give more reasonable
baryonic saturation density.

\section{Description of Nuclear Matter}\label{sec:4}

\begin{figure}[htbp]
\includegraphics[width=12.0cm]{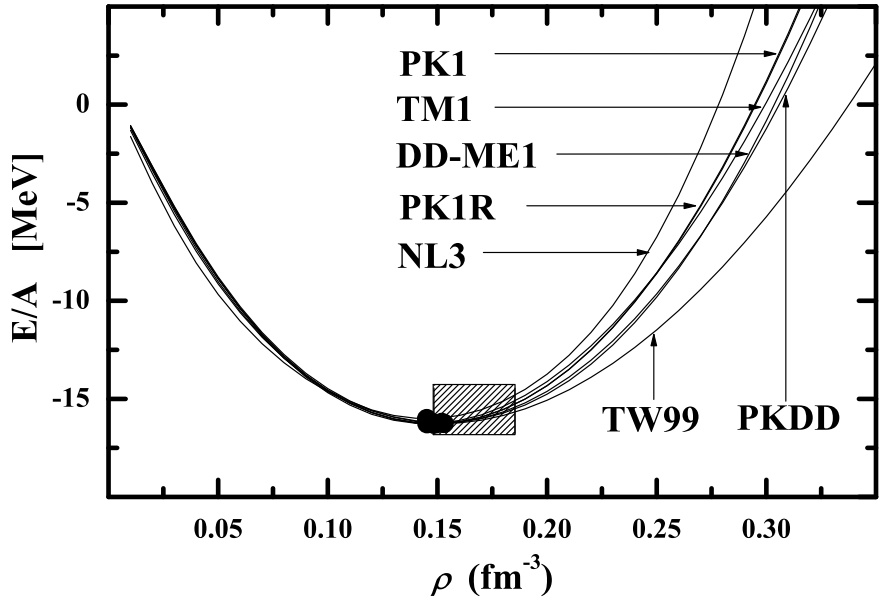}
\caption{The binding energy per particle $E/A$ in nuclear matter
as function of the baryonic density $\rho$, calculated with the
density-dependent meson-nucleon coupling effective interactions
PKDD, TW99, DD-ME1 and the nonlinear effective interactions PK1,
PK1R, TM1, NL3. The shaded area indicates the empirical value and
the filled circles represent corresponding saturation points. }
\label{fig:Eb&R}
\end{figure}

We first discuss properties of nuclear matter obtained with PK1,
PK1R and PKDD.  In Table~\ref{tab:nm-all} we compare the bulk
properties of nuclear matter with the results calculated using
TM1, NL3, TW99 and DD-ME1. In Fig.~\ref{fig:Eb&R}, the behavior of
the binding energy per particle $E/A$ as a function of the
baryonic density $\rho$ is shown. We can see that all
density-dependent meson-nucleon coupling effective interactions
give softer results than the nonlinear self-coupling ones. The
behaviors predicted by PK1, PK1R are much softer than by NL3 and a
little harder than by TM1. The results from PKDD are slightly
softer than that from DD-ME1 and much harder than that from TW99
at high density. All these behaviors can be explained in the
density-dependent meson-nucleon coupling framework.

\begin{figure}[htbp]
\includegraphics[width=12.0cm]{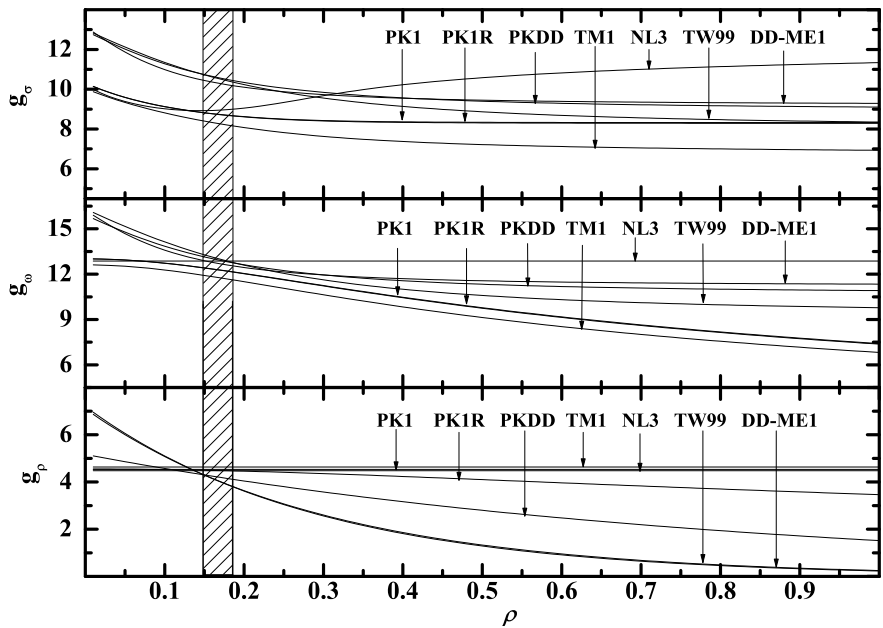}
\caption{The meson-nucleon coupling constants as a function of
baryonic density in nuclear matter. The density-dependence of
$g_\rho$ for PK1R corresponds to non-symmetric nuclear matter (N/Z
= 3).} \label{fig:gsig&gome}
\end{figure}

\begin{figure}[htbp]
\includegraphics[width=12.0cm]{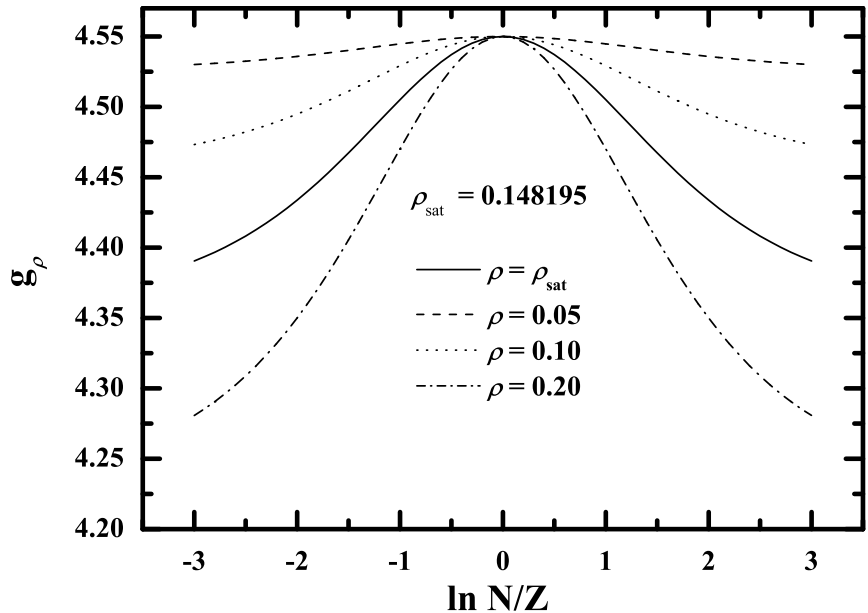}
\caption{The density-dependence of $g_\rho$ for PK1R with respect
to the neutron-proton ratio N/Z.} \label{fig:grho&qnz}
\end{figure}

As we have mentioned in expressions (\ref{couplings}), the
meson-nucleon coupling constants in the nonlinear self-coupling of
mesons can be expressed as some kind of density-dependence. Fig.
\ref{fig:gsig&gome} shows this density-dependence for the
nonlinear self-coupling effective interactions and the results for
the density-dependent version are also given for comparison. We
can see that almost all the density-dependent coupling constants
decrease with increasing density except for $g_\sigma$ of NL3,
which has a strong $\sigma$ self-coupling($g_3=-28.8851$). On the
other hand, the coupling constants $g_\sigma$ and $g_\omega$ of
TM1, which has relatively weak $\sigma$ self-coupling ($g_3 =
0.6183$) and strong $\omega$ self-coupling ($c_3 = 71.3075$), are
smaller than the others, which means that TM1 provides relatively
weaker scalar and vector potentials. This is the reason why TM1
presents the softer behavior than other nonlinear self-coupling
effective interactions. In Fig.~\ref{fig:Eb&R}, TW99 predicts the
softest results because of its relatively small $g_\omega$ as
compared with DD-ME1, PKDD and NL3, large $g_\sigma$ as compared
with PK1, PK1R and TM1 in Fig.~\ref{fig:gsig&gome}. As we know,
the repulsive potential would be dominant at high density. In
Fig.~\ref{fig:Eb&R}, NL3 gives the hardest results because of its
constant and large $g_\omega$ even though its $\sigma$-N coupling
constant $g_\sigma$ increase with the density. For the new
parameter sets PK1, PK1R and PKDD, which present the mid soft
behaviors in Fig.\ref{fig:Eb&R}, the coupling constants also lie
between the largest and the smallest in Fig. \ref{fig:gsig&gome}.
For the parameter set PK1R, the density-dependence of the $g_\rho$
is fairly weak as compared with that of the density-dependent
meson-nucleon coupling effective interactions. It can be explained
by a very weak $\rho$-field, which generate neutron-proton
symmetry field. The behavior of $g_\rho$ with respect to the
neutron-proton ratio is shown in Fig.\ref{fig:grho&qnz}. As one
can expect, the behavior is symmetric with respect to $\ln(N/Z)$
and the density-dependence becomes more remarkable with the
increase of the baryonic density and the neutron-proton asymmetry.

\section{Description of Spherical Nuclei}\label{sec:5}

\subsection{Binding Energy and Two-neutron Separation Energy}

\begin{figure}[htbp]
\includegraphics[width = 12.0cm]{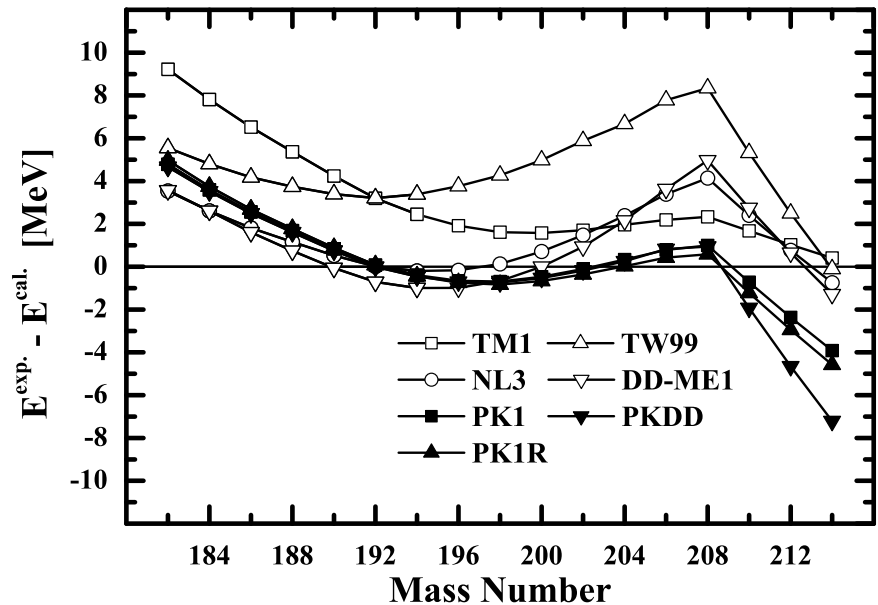}
\caption{ The deviation of the theoretical binding energies of Pb
isotopes, calculated with the nonlinear effective interactions
PK1, PK1R and density-dependent meson-nucleon coupling effective
interactions PKDD from the experimental values\cite{Audi:1995}.
The results calculated in TM1, NL3 and TW99, DD-ME1 are shown for
comparison. }\label{fig:ebn-Pb}
\end{figure}

\begin{figure}[htbp]
\includegraphics[width = 12.0cm]{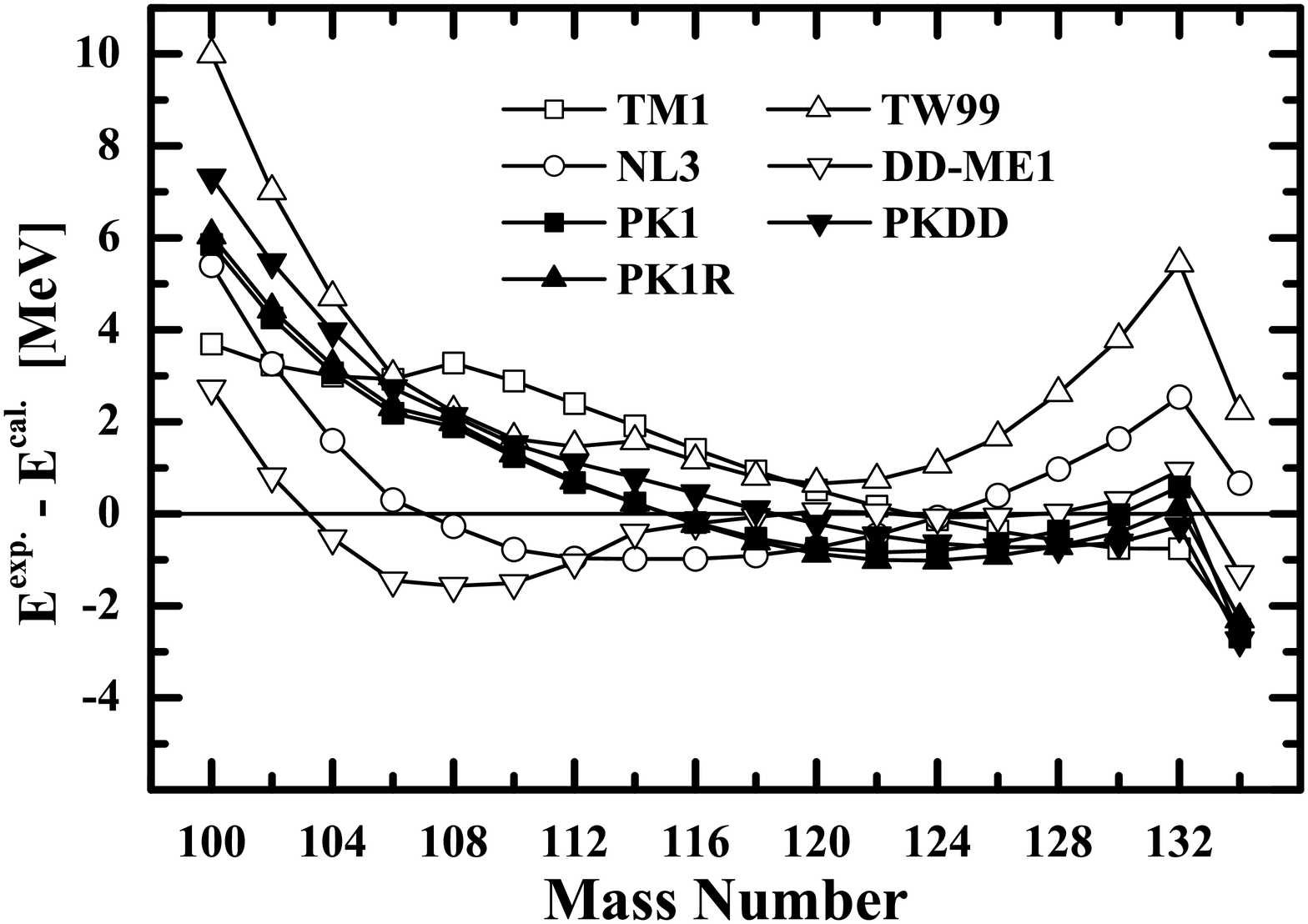}
\caption{Same as Fig. \ref{fig:ebn-Pb}, for Sn
isotopes.}\label{fig:ebn-Sn}
\end{figure}

We calculate the even-even nuclei of Pb and Sn isotopic chains
with the newly obtained parameter sets. In Table~\ref{tab:Ebn-Pb},
we compare the masses calculated with PK1, PK1R, and PKDD with the
other effective interactions and with experimental values
~\cite{Audi:1995}. Shown in Fig.~\ref{fig:ebn-Pb} are the
deviations of the masses of Pb isotopes from the
data~\cite{Audi:1995}. The results for Sn isotopes are shown in
Fig.~\ref{fig:ebn-Sn}. In Fig.~\ref{fig:ebn-Pb} and Fig.
\ref{fig:ebn-Sn} we also give the results obtained with TM1, NL3,
TW99 and DD-ME1 for comparison. All results are calculated using
the RCHB theory~\cite{Meng:1998a} where the pairing correlations
are treated self-consistently by a zero-range $\delta$-force. The
box radius is $20\ \fm$  and the pairing strength of the
zero-range $\delta$-force is $-650\ \mev$. The microscopic
center-of-mass corrections (\ref{ecm1}) and (\ref{ecm2}) are used
in all the calculations.

As we can see in Table~\ref{tab:Ebn-Pb}, Fig.~\ref{fig:ebn-Pb} and
Fig.~\ref{fig:ebn-Sn}, the newly obtained effective interactions
PK1, PK1R and PKDD provide good descriptions on the masses of both
two isotopic chains. In Fig.~\ref{fig:ebn-Pb}, all the effective
interactions overestimate the binding energy in the beginning of
the isotopic chain. However, from $^{190}$Pb to $^{210}$Pb, the
newly obtained interactions give better descriptions than all the
others. For the Sn isotopes, the density-dependent effective
interaction DD-ME1 reproduces the data very well since 4 Sn
isotopes were used in its parametrization~\cite{Niksic:2002}. From
$^{116}$Sn to $^{132}$Sn, the new effective interactions PK1, PK1R
and PKDD slightly underestimate the binding energy (less than 1
\mev). Compared with the others, the new ones still provide better
description. There exist however systematic deviations out of the
neutron magic numbers, e.g., in $^{134}$Sn and $^{214}$Pb. For the
older effective interactions, the deviations in $^{214}$Pb are
smaller but fairly large in $^{208}$Pb.

\begin{figure}[htbp]
\includegraphics[width=12.0cm]{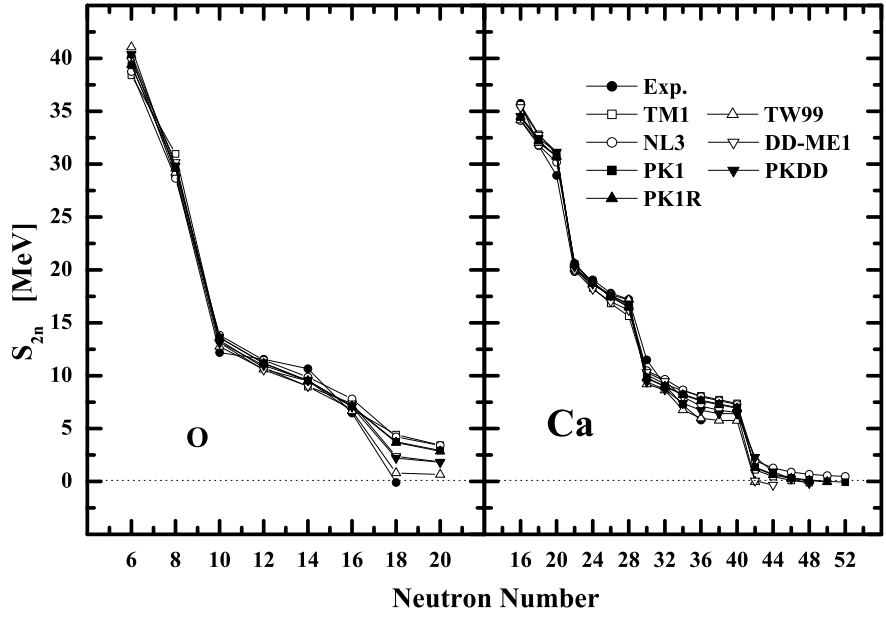}
\caption{The two-neutron separation energies in O and Ca isotopes
calculated with the nonlinear effective interactions PK1, PK1R and
the density-dependent meson-nucleon effective interaction PKDD, in
comparison with those of TM1, NL3, TW99 and DD-ME1 and the
experimental data\cite{Audi:1995}. }\label{fig:S2nO&Ca}
\end{figure}

\begin{figure}[htbp]
\includegraphics[width=12.0cm]{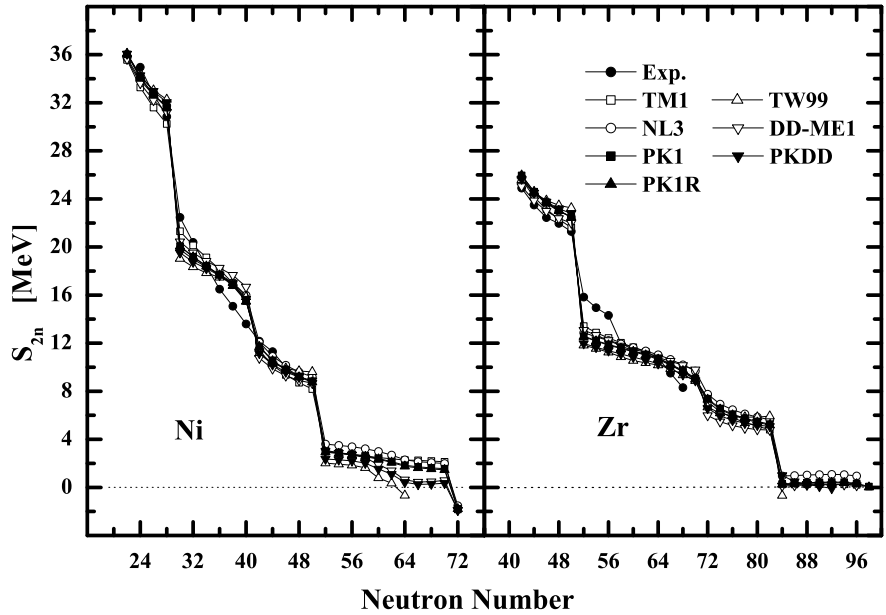}
\caption{Same as Fig. \ref{fig:S2nO&Ca}, for Ni and Zr isotopes.
}\label{fig:S2nNi&Zr}
\end{figure}

\begin{figure}[htbp]
\includegraphics[width=12.0cm]{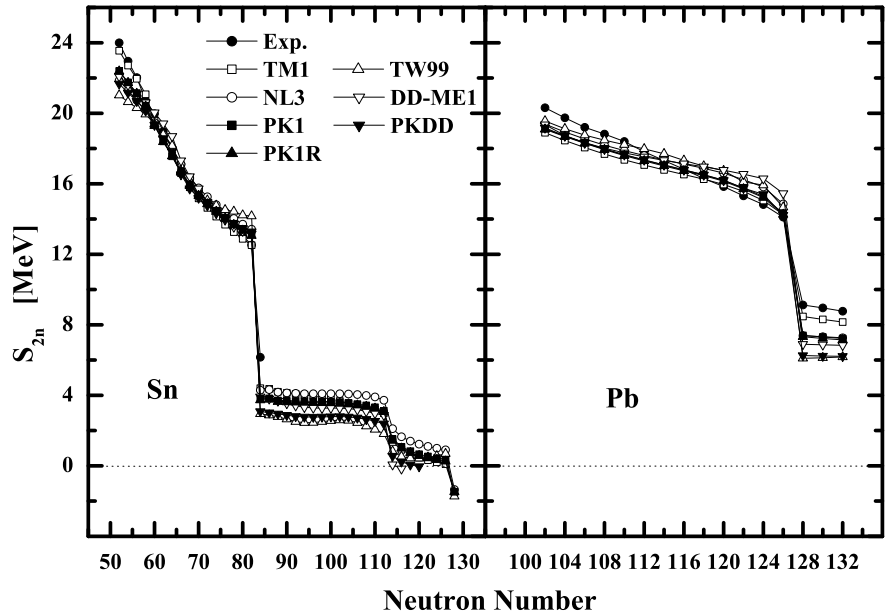}
\caption{Same as Fig. \ref{fig:S2nO&Ca}, for Sn and Pb isotopes.
}\label{fig:S2nSn&Pb}
\end{figure}

\begin{figure}[htbp]
\includegraphics[width=12.0cm]{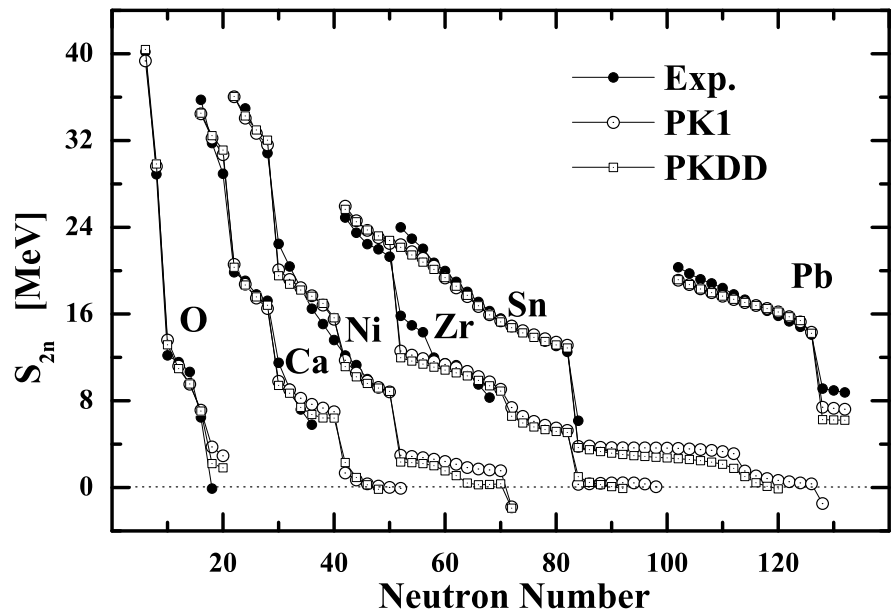}
\caption{The two-neutron separation energies calculated with the
nonlinear effective interaction PK1 and the density-dependent
meson-nucleon coupling one PKDD, as a function of the neutron
number.}\label{fig:S2n&all}
\end{figure}

From the binding energies we can extract the systematics in
two-neutron separation energies $S_{2n} = E_B(N,Z) - E_B(N-2,Z)$.
Figs.~\ref{fig:S2nO&Ca}-\ref{fig:S2nSn&Pb} exhibit two-neutron
separation energies predicted by the nonlinear self-coupling
effective interactions PK1, PK1R and the density-dependent
meson-nucleon coupling one PKDD. In comparison, the results
obtained with TM1, NL3, TW99 and DD-ME1 and the experimental
values extracted from Ref.~\cite{Audi:1995} are also given. In
Fig.~\ref{fig:S2n&all}, the systematic behaviors of two-neutron
separation energies with respect to neutron number, predicted by
the new effective interactions PK1 and PKDD, are shown. From these
figures one can see that the newly obtained interactions give a
fairly good description on the systematic behaviors in two-neutron
separation energies. In Fig.~\ref{fig:S2nNi&Zr}, we can see that
the deviations of theoretical results from experiment are rather
large for Ni isotopes with $N$ = 30-42 and for Sn isotopes with
$N$ = 52-58. Here it should be mentioned that all the theoretical
results are extracted from the calculation of the spherical RCHB
calculations~\cite{Meng:1998a} where deformation effects are not
included while they can play a significant role in these nuclei.
Furthermore, in Fig.~\ref{fig:S2n&all} we can see some unusual
phenomena along the neutron drip line which will be discussed in
the next section.

From the plots for two-neutron separation energies
(Figs.~\ref{fig:S2nO&Ca}-\ref{fig:S2n&all}), the position of the
neutron-drip line for each element seems to be determined
delicately. For the new effective interaction PK1, it predicts the
neutron-drip number $N$ = 50 for Ca, 70 for Ni, 96 for Zr and 126
for Sn. In general, the density-dependent meson-nucleon coupling
effective interaction PKDD predicts smaller neutron number of the
neutron-drip nucleus except for Ni. This may be due to its fairly
small effective mass (see Table~\ref{tab:nm-all}), which reduces
the strong attractive potential in the core and makes the coupling
between the core and valence orbital weaker. As for the
density-dependent effective interactions, TW99 predicts smaller
neutron numbers of neutron-drip nucleus for Ca, Ni and Zr while
DD-ME1 gives smaller ones for Ca and Sn. Another reason for the
deviations among the effective interactions is that the
density-dependent effective interactions give relatively large
$\rho$-N coupling at lower densities (see Fig.
\ref{fig:gsig&gome}).

\subsection{Charge Radius and Isotope Shift in Pb isotopes}
\begin{figure}[htbp]
\includegraphics[width = 12.0cm]{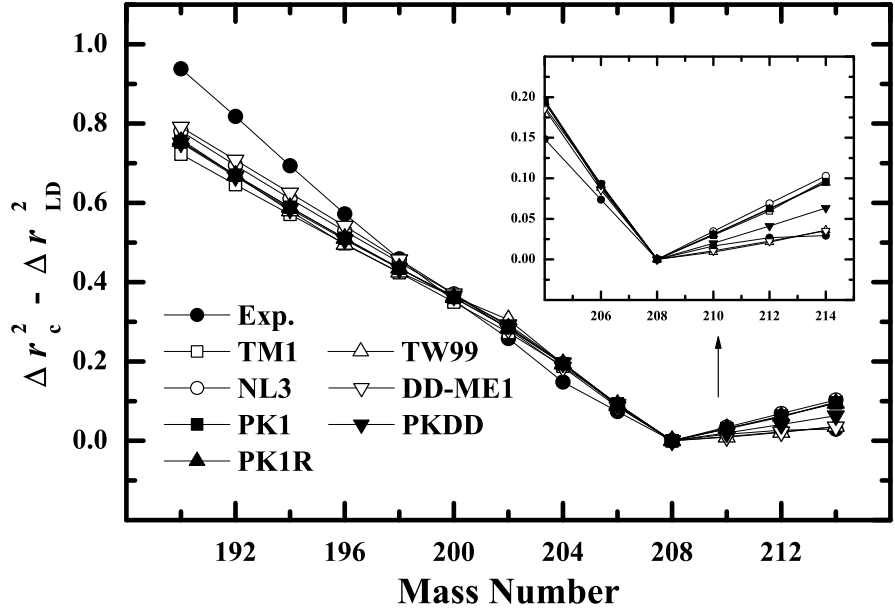}
\caption{The isotope shifts of the charge radius in Pb isotopes
(in fm$^2$), calculated with  the nonlinear effective interactions
PK1, PK1R and the density-dependent meson-nucleon coupling
effective interaction PKDD, in comparison with those of TM1, NL3,
TW99, DD-ME1 and experimental values, where $\Delta r_c^2 =
r_c^2(A) - r_c^2(208)$ and $ \Delta r_{LD}^2 = r_{LD}^2(A) -
r_{LD}^2(208)$. }\label{fig:shift-Pb}
\end{figure}

Although the radii are not included in our fitting procedure, the
newly obtained effective interactions reproduce the charge radii
of stable nuclei fairly well (see Table~\ref{tab:charge-all}). The
comparison between experimental data~\cite{Fricke:1995,Dutta:1991}
and theoretical results for the charge radii (\ref{rch}) of Pb
isotopes are shown in Table~\ref{tab:rch-Pb}.  We can see that the
new effective interactions PK1, PK1R and PKDD reproduce better the
experimental values, as compared to overestimations by TM1, NL3
and DD-ME1 and underestimations by TW99. We also calculate the
isotope-shift of charge radii for Pb isotopes with these effective
interactions. The results are shown in Fig. \ref{fig:shift-Pb}.
The kink around $^{208}$Pb is well reproduced by all the
interactions. The inset of Fig.~\ref{fig:shift-Pb} shows that the
density-dependent meson-nucleon coupling effective interactions
represent more reasonable agreement with the experimental values
than the nonlinear self-coupling ones. The new parameter sets PK1
and PK1R give slightly better results than TM1 and NL3.

\begin{figure}[htbp]
\includegraphics[width=12.0cm]{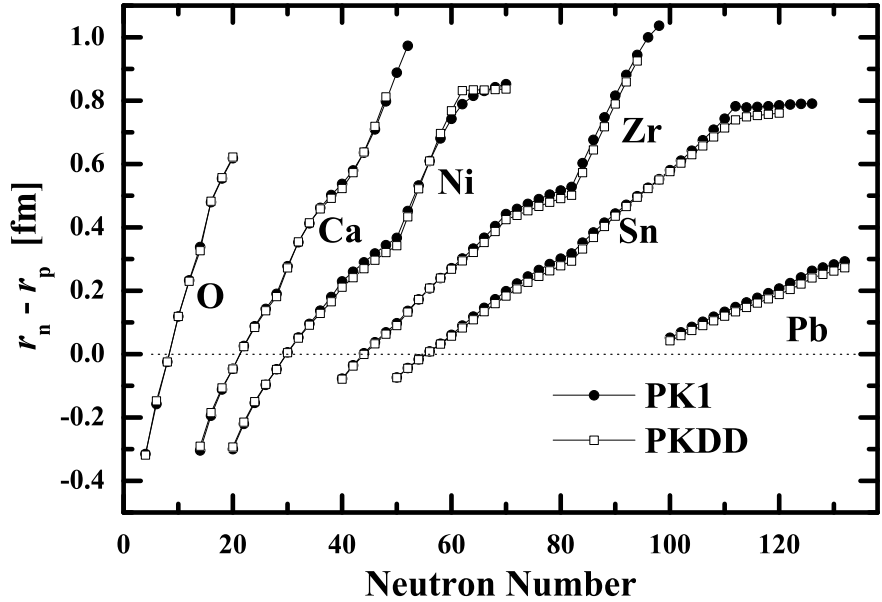}
\caption{The radius difference $r_n-r_p$ calculated with the
nonlinear effective interaction PK1 and density-dependent
meson-nucleon coupling one PKDD, with respect to the neutron
number.} \label{fig:Skin&all}
\end{figure}

Besides the charge radii, we have also used the new effective
interactions to investigate the systematic behavior of the neutron
skin (the difference between neutron and proton radii) along the
isotope chains: O, Ca, Ni, Zr, Sn and Pb. Fig.~\ref{fig:Skin&all}
shows the radii difference $r_n -r_p$ calculated with the new sets
PK1 and PKDD. As we have seen in Fig.~\ref{fig:S2n&all}, there are
several weakly-bound nuclei near the neutron drip line of Ca, Ni,
Zr and Sn, whose two-neutron separation energies stay around zero
over a range of numbers. This is in general a kind of signal for
the existence of a neutron skin or halo. However from
Fig.\ref{fig:Skin&all} one can see that the difference $r_n-r_p$
tends to be a constant near the drip line for the Ni and Sn
isotopes thus indicating a neutron skin rather than a neutron
halo. On the other hand, the results support the existence of a
neutron halo in Ca isotopes \cite{Meng:2002PRC} and a giant
neutron halo in Zr isotopes \cite{Meng:1998} because their neutron
distributions tend to be more dispersive and $r_n -r_p$ keeps
increasing rapidly.

\subsection{Single-Particle Energy and Spin-orbit Splitting}
\begin{figure}
\includegraphics[width=12.0cm]{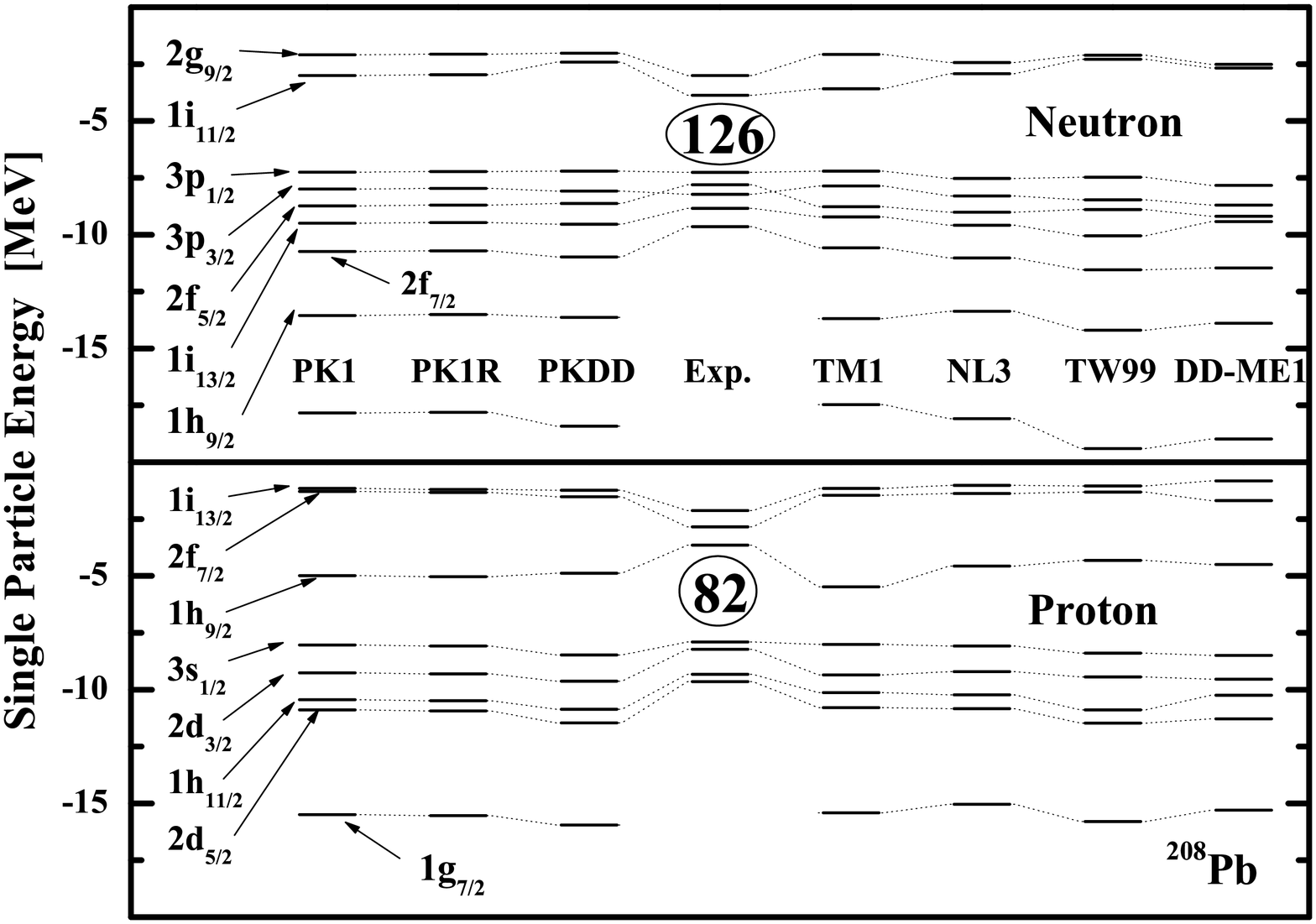}
\caption{The single-particle energies in $^{208}$Pb, calculated
with PK1, PK1R and PKDD, in comparison with the results of TM1,
NL3, TW99, DD-ME1 and experimental values.}\label{fig:Pb208lev}
\end{figure}

\begin{figure}
\includegraphics[width=12.0cm]{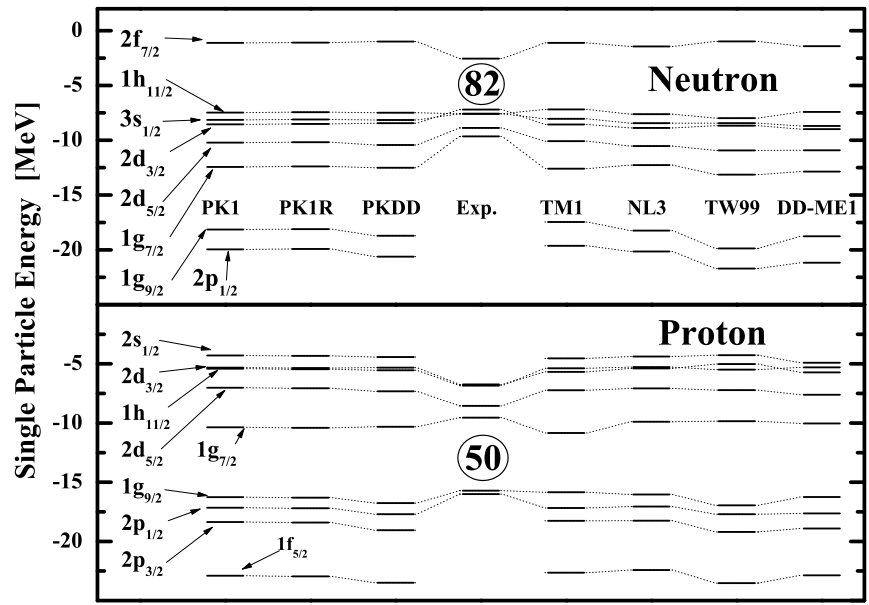}
\caption{Same as Fig. \ref{fig:Pb208lev}, for $^{132}$Sn.
}\label{fig:Sn132lev}
\end{figure}

\begin{figure}
\includegraphics[width=12.0cm]{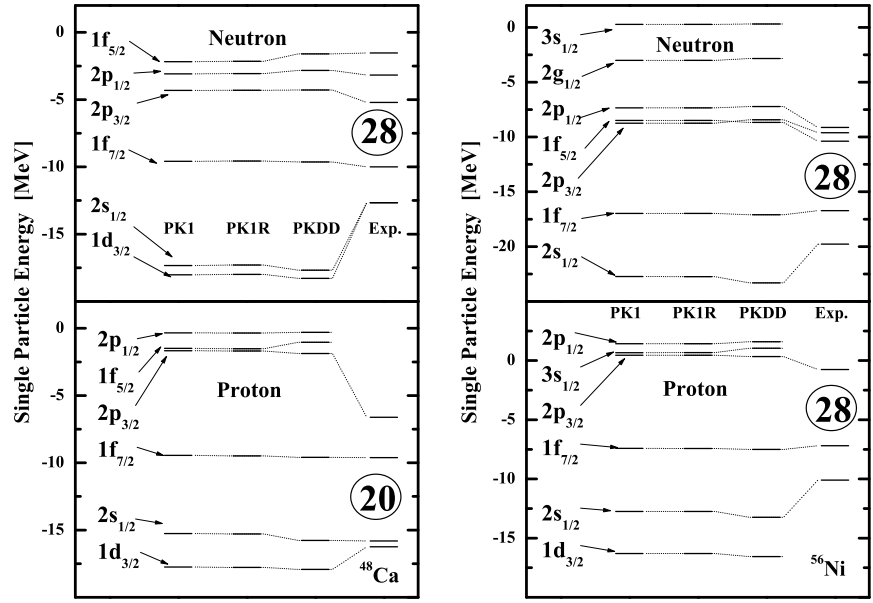}
\caption{The single-particle energies in $^{48}$Ca and $^{56}$Ni,
calculated with PK1, PK1R and PKDD, in comparison with
experimental values.}\label{fig:Ca48&Ni56}
\end{figure}

\begin{figure}[htbp]
\includegraphics[width=12.0cm]{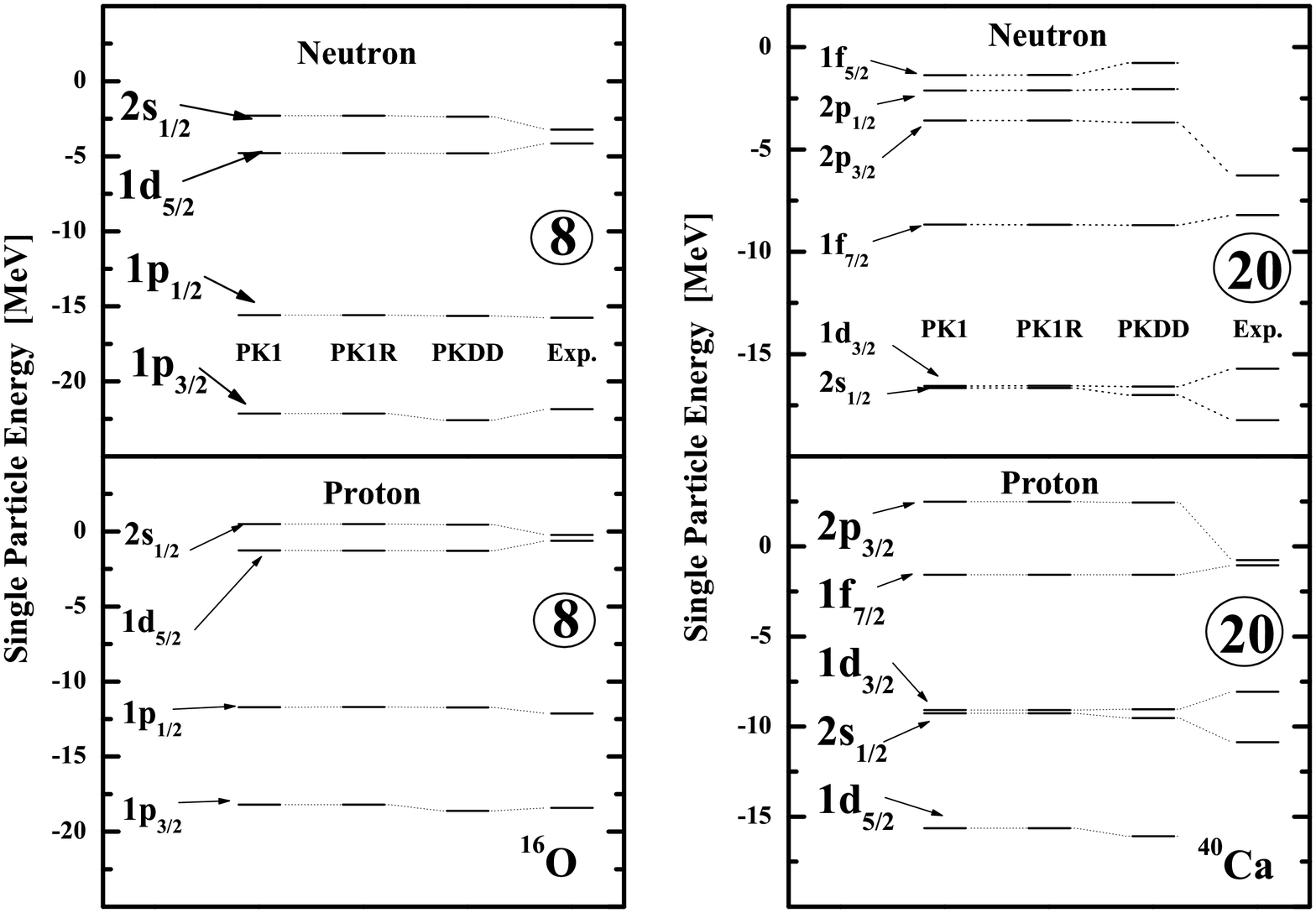}
\caption{Same as Fig.\ref{fig:Ca48&Ni56}, for $^{16}$O and
$^{40}$Ca.}\label{fig:O16&Ca40}
\end{figure}

The RMF theory is a microscopic theory with a limited number of
parameters. It can give the detailed microscopic structure of
nuclei. Figs.~\ref{fig:Pb208lev} to~\ref{fig:O16&Ca40} show the
single-particle energies of the doubly magic nuclei calculated
with the newly obtained parameter sets PK1, PK1R and PKDD. In
Figs.~\ref{fig:Pb208lev} and~\ref{fig:Sn132lev} the results
obtained with TM1, NL3, TW99 and DD-ME1 are also given for
comparison.  As we know,  it is not straightforward to compare
with the experimental results away from Fermi energies because of
dynamical coupling not included in RMF. Here, the experimental
values are extracted from one-nucleon separation
energies~\cite{Audi:1995} and resonance energies~\cite{nudat}. In
these plots, we find systematic agreements with the experimental
results. The single-particle energies near the magic numbers agree
well with the experimental values. For levels far away from the
Fermi energies systematic deviations appear. The states below the
Fermi energy seems to be too strongly bound, whereas the states
above the Fermi energy show underbinding as compared to the
experimental values. In fact, the experimental resonances are not
simply single particle states. The coupling with core excitations
through the residual interaction, which is obviously not included
in the mean field model, leads in general to a shift of the
resonance energies in the direction to the Fermi energy. In these
plots of single-particle energies, the ordering of levels is well
described by the new effective interactions except for somewhat
different ordering of neutron levels in $^{132}$Sn and $^{208}$Pb,
a common feature of all the effective interactions.

From the single-particle energies mentioned above, we can extract
the spin-orbit splittings. Table~\ref{tab:splitting} shows the
spin-orbit splitting calculated with the newly obtained effective
interactions in the doubly magic nuclei. The experimental values
and the results calculated with other interactions are also given
for comparison in Table~\ref{tab:splitting}. The new effective
interactions reproduce well the spin-orbit splittings. For the
density-dependent meson-nucleon coupling effective interaction
PKDD, the spin-orbit splitting turns out to be larger than the
experimental values (except for the neutron 2p in $^{48}$Ca and
the neutron 3p in $^{208}$Pb). The nonlinear self-coupling
effective interactions PK1 and PK1R predict smaller splitting and
improve the agreement with the experiment. The behavior can be
explained by the relatively large self-energies predicted in our
parametrization, which determine the strength of spin-orbit
splitting. As shown in Table~\ref{tab:nm-all}, the nonlinear
self-coupling effective interactions predict a larger effective
mass than the density-dependent meson-nucleon coupling ones: the
smaller the effective mass, the larger the spin-orbit splitting.
This is  also the reason why TM1 gives the smallest splitting and
TW99 the largest.

\section{Summary}\label{sec:6}

In this work, we have searched for new effective interactions to
describe both stable and unstable nuclei in the relativistic mean
field (RMF) theory with nonlinear self-coupling or
density-dependent meson-nucleon coupling. In order to give a more
precise description on the mass of nuclei, the microscopic
center-of-mass correction is introduced, which makes it possible
to give a unified description with one effective interaction for
the nuclei from the light area to heavy area. As an elicitation
from the density-dependent meson-nucleon coupling RMF theory, we
introduce the nonlinear self-coupling for $\rho$-field.
We obtain three new effective interactions: PK1 with nonlinear
self-coupling of $\sigma$-field and $\omega$-field, PK1R with
nonlinear self-coupling of $\sigma$-field, $\omega$-field and
$\rho$-field and PKDD with density-dependent meson-nucleon
coupling.

With the newly obtained parameter sets, we investigate the
behavior of the binding energy per particle and the meson-nucleon
coupling constants with respect to the baryonic density in nuclear
matter. The new sets PK1, PK1R and PKDD provide an appropriate
description. Compared with TM1 and NL3, the new ones give a more
reasonable baryonic saturation density.

We then calculated the usual reference nuclei and Pb, Sn isotope
chains and compared the masses with the available data. The
nonlinear self-coupling effective interactions PK1, PK1R and
density-dependent meson-nucleon coupling effective interaction
PKDD reproduce well the data. As compared with other existing
effective interactions, the new ones also provide a good
description of the charge radii of the usual stable nuclei and Pb
isotopes.

We have also studied the systematics of two-neutron separation
energies and neutron skin in isotopic chains. The two-neutron
separation energies provided by the new  interactions PK1, PK1R
and PKDD are in good agreement with the experiment. We have also
investigated the behavior of two-neutron separation energies and
neutron skin near the neutron-drip line and given a reasonable
interpretation of the formation of neutron halos.

The single-particle energies and spin-orbit splittings in doubly
magic nuclei predicted by the new parameter sets are compared with
the experimental values and with other effective interactions. The
new effective interactions PK1, PK1R and PKDD give a reasonable
description of spin-orbit splittings and single-particle energies
as compared with the experimental results. The systematic behavior
of the spin-orbit splitting is interpreted in comparison with
other effective interactions.

Combining with the above information, we come to a conclusion that
the new parameter sets PK1, PK1R and PKDD give better description
for finite nuclei than other effective interactions.

\begin{acknowledgments}
This work is partly supported by the Major State Basic
Research Development Program Under Contract Number G2000077407 and
the National Natural Science Foundation of China under Grant No.
10025522, 10047001, 10221003 and 19935030.
\end{acknowledgments}

\begin{table}[htbp]
\caption{ The nonlinear effective interactions PK1, PK1R and
density-dependent effective interactions PKDD. The masses (in MeV)
and meson-nucleon couplings are shown in comparison with
TM1~\cite{Sugahara:1994}, NL3~\cite{Lalazissis:1997} and
TW99~\cite{Typel:1999} and DD-ME1~\cite{Niksic:2002}}
\label{tab:psets} \brule \btab{cccccccc}
&PK1&PK1R&PKDD&TM1&NL3&TW99&DD-ME1\\ \hline
$M_n$&939.5731&939.5731&939.5731&938&939&939&938.5000\\
$M_p$&938.2796&938.2796&938.2796&938&939&939&938.5000\\
$m_\sigma$&514.0891&514.0873&555.5112&511.198&508.1941&550&549.5255\\
$m_\omega$&784.254&784.2223&783&783&782.501&783&783.0000\\
$m_\rho$&763&763&763&770&763&763&763.0000\\
$g_\sigma$&10.3222&10.3219&10.7385&10.0289&10.2169&10.7285&10.4434\\
$g_\omega$&13.0131&13.0134&13.1476&12.6139&12.8675&13.2902&12.8939\\
$g_\rho$&4.5297&4.55&4.2998&4.6322&4.4744&3.661&3.8053\\
$g_2$&$-$8.1688&$-$8.1562&0&$-$7.2325&$-$10.4307&0&0.0000\\
$g_3$&$-$9.9976&$-$10.1984&0&0.6183&$-$28.8851&0&0.0000\\
$c_3$&55.636&54.4459&0&71.3075&0&0&0.0000\\
$d_3$&0&350&0&0&0&0&0 \etab

\caption{Density-Dependent parameters of PKDD for meson-nucleon
coupling in comparison with TW99~\cite{Typel:1999} and
DD-ME1~\cite{Niksic:2002}}\label{tab:pkdd} \btab{cccccccccc}
&$a_\sigma$&$b_\sigma$&$c_\sigma$&$d_\sigma$&$a_\omega$&$b_\omega$
&$c_\omega$&$d_\omega$&$a_\rho$\\
\hline
PKDD&1.327423&0.435126&0.691666&0.694210&1.342170&0.371167&0.611397&0.738376&0.183305\\
TW99&1.365469&0.226061&0.409704&0.901995&1.402488&0.172577&0.344293&0.983955&0.515000\\
DD-ME1&1.3854&0.9781&1.5342&0.4661&1.3879&0.8525&1.3566&0.4957&0.5008\\
\etab \erule
\end{table}

\begin{table}[htbp]
\caption{Total binding energies (in MeV) calculated with the
nonlinear effective interactions PK1, PK1R and density-dependent
meson-nucleon coupling effective interaction PKDD are shown in
comparison with those of TM1~\cite{Sugahara:1994},
NL3~\cite{Lalazissis:1997} , TW99~\cite{Typel:1999},
 DD-ME1~\cite{Niksic:2002}
and experimental  data~\cite{Audi:1995}. The bold-faced quantities
denote the observables used in the
parametrizations.}\label{tab:ebn-all} \brule \btab{ccccccccccc}
Nucleus&Exp. &  PK1& PK1R  &PKDD&TM1& NL3& TW99&    DD-ME1\\
\hline $^{16}$O
&$-$127.619&\bf{$-$128.094}&\bf{$-$128.047}&\bf{$-$127.808}&$-$128.951&\bf{$-$127.127}
&\bf{$-$128.147}&\bf{$-$127.926}\\
$^{24}$O  &$-$168.500&$-$169.558&$-$169.381&$-$168.542&$-$168.858&$-$170.116&\bf-167.693&$-$167.949 \\
$^{40}$Ca
&$-$342.052&\bf{$-$342.773}&\bf-342.741&\bf-342.579&\bf-344.661&\bf-341.709&\bf-343.352
&\bf-343.653\\
$^{48}$Ca &$-$415.991&\bf-416.077&\bf$-$415.974&\bf$-$415.944&\bf$-$415.668&\bf$-$415.377&\bf$-$416.888&\bf$-$415.012\\
$^{56}$Ni &$-$483.998&\bf$-$483.956&\bf$-$484.031&\bf$-$484.479&$-$480.620&$-$483.599&\bf$-$487.096&$-$480.869\\
$^{58}$Ni &$-$506.454&$-$504.033&$-$504.091&$-$504.013&\bf$-$501.933&\bf$-$503.395&$-$506.128&$-$501.312\\
$^{68}$Ni &$-$590.430&\bf$-$591.685&\bf$-$591.559&\bf$-$591.241&$-$591.845&$-$591.456&$-$592.676&$-$592.253\\
$^{90}$Zr &$-$783.893&\bf$-$784.781&\bf$-$784.788&\bf$-$784.879&\bf$-$785.281&\bf$-$783.859&\bf$-$786.625&\bf$-$784.206\\
$^{112}$Sn&$-$953.529&$-$954.210&$-$954.251&$-$953.730&$-$955.925&$-$952.562&$-$954.991&\bf$-$952.468\\
$^{116}$Sn&$-$988.681&\bf$-$988.491&\bf$-$988.460&\bf$-$988.066&\bf$-$990.083&\bf$-$987.699&$-$989.842&\bf$-$988.470\\
$^{124}$Sn&$-$1049.963&$-$1049.162&$-$1048.948&$-$1048.113&\bf$-$1049.832&\bf$-$1049.884&$-$1051.033&\bf$-$1049.880\\
$^{132}$Sn&$-$1102.920&\bf$-$1103.503&\bf$-$1103.053&\bf$-$1102.648&$-$1102.163&\bf$-$1105.459&$-$1108.363
&\bf$-$1103.857\\
$^{184}$Pb&$-$1431.960&$-$1435.548&$-$1435.706&$-$1435.477&\bf$-$1439.768&$-$1434.569&$-$1436.761&$-$1434.569\\
$^{194}$Pb&$-$1525.930&\bf$-$1525.536&\bf$-$1525.494&\bf$-$1525.474&$-$1528.378&$-$1525.733&$-$1529.309&$-$1524.937\\
$^{196}$Pb&$-$1543.250&$-$1542.592&$-$1542.502&$-$1542.545&\bf$-$1545.163&$-$1543.085&$-$1547.012&$-$1542.262\\
$^{204}$Pb&$-$1607.520&$-$1607.851&$-$1607.545&$-$1607.770&$-$1609.477&$-$1609.906&$-$1608.246&\bf$-$1609.676\\
$^{208}$Pb&$-$1636.446&\bf$-$1637.443&\bf$-$1637.024&\bf$-$1637.387&\bf$-$1638.777&\bf$-$1640.584&
\bf$-$1644.790&\bf$-$1641.415\\
$^{214}$Pb&$-$1663.298&$-$1659.382&$-$1658.718&$-$1656.084&\bf$-$1663.706&\bf$-$1662.551&$-$1656.086&\bf$-$1662.011\\
$^{210}$Po&$-$1645.228&
$-$1648.443&$-$1648.102&$-$1648.039&$-$1650.819&$-$1650.755&$-$1654.271&\bf$-$1651.482\\
\hline $\Delta$\footnote{The total square deviation from the
experimental values $\Delta^2 = \sum_i \lb E^{\text{exp.}}_i -
E^{\text{cal.}}_i\rb^2$}
&&7.1980&7.4207&9.2744&12.8135&9.1140&17.6762&11.1580\\
$\delta$\footnote{The relative square deviation $\delta^2 = \sum_i
\lb E^{\text{exp.}}_i -
E^{\text{cal.}}_i\rb^2/E^{\text{exp.}2}_i$}
&&0.0102&0.0094&0.0080&0.0192&0.0135&0.0159&0.0152\\
\etab \erule
\end{table}

\begin{table}[htbp]
\caption{Charge radii (in fm) calculated with the nonlinear
effective interactions PK1, PK1R and density-dependent
meson-nucleon coupling effective interaction PKDD are shown in
comparison with those of TM1~\cite{Sugahara:1994},
NL3~\cite{Lalazissis:1997} , TW99~\cite{Typel:1999},
DD-ME1~\cite{Niksic:2002} and experimental
data~\cite{Fricke:1995}.}\label{tab:charge-all}
\brule\btab{ccccccccccc} Nucleus&Exp. &  PK1& PK1R  &PKDD&TM1&
NL3& TW99&    DD-ME1\\ \hline
$^{16}$O&2.693&2.6957&2.6959&2.6988&2.7026&2.7251&2.6799&2.7268\\
$^{24}$O&&2.8106&2.8108&2.8184&2.8364&2.8286&2.8049&2.8543\\
$^{40}$Ca& 3.478&3.4433&3.4435&3.4418&3.4541&3.4679&3.4151&3.4622\\
$^{48}$Ca& 3.479&3.4675&3.4675&3.4716&3.4911&3.4846&3.4510&3.4946\\
$^{56}$Ni&&3.7085&3.7084&3.7162&3.7471&3.7122&3.6867&3.7315\\
$^{58}$Ni& 3.776&3.7383&3.7381&3.7442&3.7755&3.7435&3.7158&3.7613\\
$^{68}$Ni&&3.8621&3.8620&3.8681&3.8901&3.8773&3.8491&3.8926\\
$^{90}$Zr& 4.270&4.2522&4.2521&4.2534&4.2799&4.2689&4.2278&4.2725\\
$^{112}$Sn&4.593&4.5704&4.5701&4.5722&4.6021&4.5861&4.5461&4.5901\\
$^{116}$Sn&4.625&4.5984&4.5981&4.6004&4.6303&4.6149&4.5758&4.6212\\
$^{124}$Sn&4.677&4.6536&4.6533&4.6567&4.6874&4.6685&4.6331&4.6781\\
$^{132}$Sn&& 4.7064&4.7061&4.7102&4.7442&4.7183&4.6842&4.7270\\
$^{184}$Pb&& 5.3806&5.3801&5.3807&5.4156&5.3996&5.3511&5.4002\\
$^{194}$Pb&5.442&5.4327&5.4322&5.4329&5.4712&5.4506&5.4017&5.4539\\
$^{196}$Pb&5.449&5.4438&5.4433&5.4440&5.4826&5.4614&5.4123&5.4645\\
$^{204}$Pb&5.482&5.4869&5.4864&5.4877&5.5261&5.5027&5.4837&5.5038\\
$^{208}$Pb&5.504&5.5048&5.5043&5.5053&5.5444&5.5204&5.4750&5.5224\\
$^{214}$Pb&5.559&5.5658&5.5653&5.5635&5.6052&5.5820&5.5603&5.5779\\
$^{210}$Po&&
5.5370&5.5365&5.5371&5.5762&5.5539&5.5070&5.5544\\\hline
$\Delta$\footnote{The total square deviation from the experimental
values $\Delta^2 =\sum_i\lb
r_i^{\text{exp.}}-r_i^{\text{cal.}}\rb^2$}
&&0.0708&0.0712&0.0655&0.0941&0.0625&0.1436&0.0588\\
$\delta$\footnote{The relative square deviation $\delta^2 = \sum_i
\lb r_i^{\text{exp.}}-r_i^{\text{cal.}}\rb^2/r_i^{\text{exp.}2}$}
&&0.0178&0.0179&0.0166&0.0185&0.0169&0.0346&0.0163\\
\etab\erule
\end{table}

\begin{table}[htbp]
\caption{Nuclear matter properties calculated with the nonlinear
effective interactions PK1, PK1R and the density-dependent
effective interactions PKDD are shown in comparison with
TM1~\cite{Sugahara:1994}, NL3~\cite{Lalazissis:1997},
TW99~\cite{Typel:1999},
DD-ME1~\cite{Niksic:2002}.}\label{tab:nm-all} \brule
\btab{lcccccc} Interaction& $\rho_{\text{sat.}}(\fm^{-3})$&$E_b$
[MeV]&K [MeV]&$J$ [MeV]& $M^*/M$(n)& $M^*/M$(p)\\ \hline
PK1    &0.148195 &$-$16.268    &282.644   &37.641    &0.605525    &0.604981\\
PK1R   &0.148196 &$-$16.274    &283.674   &37.831    &0.605164    &0.604620\\
PKDD   &0.149552 &$-$16.267    &262.181   &36.790    &0.571156    &0.570565\\
NL3    &0.145115 &$-$16.005    &267.998   &36.558    &0.603761    &0.603761\\
TM1    &0.145218 &$-$16.263    &281.161   &36.892    &0.634395    &0.634395\\
TW99   &0.153004 &$-$16.247    &240.276   &32.767    &0.554913    &0.554913\\
DD-ME1 &0.151962 &$-$16.201    &244.719   &33.065    &0.577960    &0.577960\\
\etab \erule
\end{table}

\begin{table}[htbp]
\caption{Total binding energies of Pb isotopes (in MeV) calculated
with the nonlinear self-coupling effective interactions PK1, PK1R
and the density-dependent meson-nucleon coupling effective
interaction PKDD, in comparison with the experimental
values~\cite{Audi:1995} and the results of TM1, NL3, TW99 and
DD-ME1.}\label{tab:Ebn-Pb}
\brule \btab{ccccccccc} A&Exp.&PK1&PK1R&PKDD&TM1&NL3&TW99&DD-ME1\\
\hline
182&$-$1411.650&$-$1416.431&$-$1416.619&$-$1416.309&$-$1420.872&$-$1415.184&$-$1417.202&$-$1415.216\\
184&$-$1431.960&$-$1435.548&$-$1435.706&$-$1435.477&$-$1439.768&$-$1434.569&$-$1436.761&$-$1434.569\\
186&$-$1451.700&$-$1454.258&$-$1454.382&$-$1454.192&$-$1458.222&$-$1453.511&$-$1455.879&$-$1453.306\\
188&$-$1470.900&$-$1472.586&$-$1472.673&$-$1472.504&$-$1476.272&$-$1472.056&$-$1474.640&$-$1471.639\\
190&$-$1489.720&$-$1490.555&$-$1490.603&$-$1490.468&$-$1493.958&$-$1490.247&$-$1493.109&$-$1489.657\\
192&$-$1508.120&$-$1508.197&$-$1508.201&$-$1508.116&$-$1511.317&$-$1508.130&$-$1511.329&$-$1507.411\\
194&$-$1525.930&$-$1525.536&$-$1525.494&$-$1525.474&$-$1528.378&$-$1525.733&$-$1529.309&$-$1524.937\\
196&$-$1543.250&$-$1542.592&$-$1542.502&$-$1542.545&$-$1545.163&$-$1543.085&$-$1547.012&$-$1542.262\\
198&$-$1560.070&$-$1559.378&$-$1559.236&$-$1559.329&$-$1561.685&$-$1560.199&$-$1564.338&$-$1559.403\\
200&$-$1576.365&$-$1575.893&$-$1575.697&$-$1575.831&$-$1577.942&$-$1577.076&$-$1581.344&$-$1576.365\\
202&$-$1592.202&$-$1592.095&$-$1591.843&$-$1592.022&$-$1593.905&$-$1593.679&$-$1598.084&$-$1593.133\\
204&$-$1607.520&$-$1607.851&$-$1607.545&$-$1607.770&$-$1609.477&$-$1609.906&$-$1614.197&$-$1609.676\\
206&$-$1622.340&$-$1623.126&$-$1622.765&$-$1623.167&$-$1624.530&$-$1625.725&$-$1630.122&$-$1625.966\\
208&$-$1636.446&$-$1637.443&$-$1637.024&$-$1637.387&$-$1638.777&$-$1640.584&$-$1644.790&$-$1641.415\\
210&$-$1645.568&$-$1644.844&$-$1644.345&$-$1643.643&$-$1647.245&$-$1647.969&$-$1650.888&$-$1648.312\\
212&$-$1654.524&$-$1652.155&$-$1651.573&$-$1649.873&$-$1655.549&$-$1655.289&$-$1657.020&$-$1655.174\\
214&$-$1663.298&$-$1659.382&$-$1658.718&$-$1656.084&$-$1663.706&$-$1662.551&$-$1663.196&$-$1662.011\\
\etab\erule
\end{table}

\begin{table}[htbp]
\caption{Charge radii of Pb isotopes (in fm), calculated with  the
nonlinear effective interactions PK1, PK1R and the
density-dependent meson-nucleon coupling effective interaction
PKDD, in comparison with those of TM1, NL3, TW99, DD-ME1 and
experimental
values~\cite{Fricke:1995,Dutta:1991}.}\label{tab:rch-Pb} \brule
\btab{ccccccccc} A&Exp.&PK1&PK1R&PKDD&TM1&NL3&TW99&DD-ME1\\ \hline
190&5.4273&5.4112&5.4107&5.4115&5.4486&5.4295&5.3814&5.4325\\
192&5.4347&5.4219&5.4214&5.4221&5.4599&5.4400&5.3915&5.4433\\
194&5.4416&5.4327&5.4322&5.4329&5.4712&5.4506&5.4017&5.4539\\
196&5.4487&5.4438&5.4433&5.4440&5.4826&5.4614&5.4123&5.4645\\
198&5.4564&5.4551&5.4546&5.4555&5.4940&5.4722&5.4241&5.4748\\
200&5.4649&5.4664&5.4659&5.4673&5.5052&5.4830&5.4366&5.4850\\
202&5.4741&5.4773&5.4768&5.4786&5.5161&5.4934&5.4493&5.4947\\
204&5.4820&5.4869&5.4864&5.4877&5.5261&5.5027&5.4571&5.5038\\
206&5.4930&5.4954&5.4949&5.4960&5.5352&5.5109&5.4653&5.5125\\
208&5.5040&5.5048&5.5043&5.5053&5.5444&5.5204&5.4750&5.5224\\
210&5.5231&5.5252&5.5247&5.5247&5.5645&5.5411&5.4935&5.5409\\
212&5.5415&5.5455&5.5450&5.5441&5.5846&5.5616&5.5122&5.5594\\
214&5.5591&5.5658&5.5653&5.5635&5.6052&5.5820&5.5310&5.5779\\
\etab \erule
\end{table}

\begin{table}[htbp]
\caption{The theoretical (calculated with the nonlinear effective
interactions PK1, PK1R, TM1~\cite{Sugahara:1994},
NL3~\cite{Lalazissis:1997} and the density-dependent ones PKDD,
TW99~\cite{Typel:1999}, DD-ME1~\cite{Niksic:2002}) and
experimental spin-orbit splittings (in MeV) of neutron ($\nu$) and
proton ($\pi$) levels in doubly magic
nuclei.}\label{tab:splitting} \brule\btab{cccccccccc}
Nucleus&State&PK1&PK1R&PKDD&Exp.&TM1&NL3&TW99&DD-ME1 \\ \hline
\multirow{2}{0.5cm}{$^{16}$O}&$\nu$1p&6.550&6.550&6.950&6.180&5.660&6.480&7.480&6.320 \\
&$\pi$1p&6.490&6.500&6.900&6.320&5.610&6.400&7.410&6.250 \\ \hline
\multirow{2}{0.5cm}{$^{48}$Ca}&$\nu$1f&7.411&7.417&8.028&8.380&6.486&7.493&8.687&7.498 \\
&$\nu$2p&1.238&1.237&1.459&2.020&1.142&1.330&1.567&1.458 \\ \hline
\multirow{2}{0.5cm}{$^{56}$Ni}&$\nu$1f&8.223&8.231&8.670&7.160&6.907&8.703&9.274&8.067 \\
&$\nu$2p&1.141&1.142&1.442&1.110&1.105&1.112&1.573&1.388 \\ \hline
\multirow{3}{0.5cm}{$^{132}$Sn}&$\nu$2d&1.659&1.662&1.990&1.650&1.515&1.661&2.257&1.940 \\
&$\pi$1g&5.900&5.910&6.450&6.080&5.010&6.152&7.108&6.210 \\
&$\pi$2d&1.704&1.706&2.005&1.750&1.556&1.690&2.203&1.893 \\ \hline
\multirow{5}{0.5cm}{$^{208}$Pb}&$\nu$2f&2.005&2.008&2.356&1.770&1.812&2.011&2.648&2.268 \\
&$\nu$1i&6.492&6.503&7.126&5.840&5.634&6.665&7.761&6.748 \\
&$\nu$3p&0.742&0.742&0.879&0.900&0.657&0.764&0.992&0.866 \\
&$\pi$2d&1.626&1.622&1.832&1.330&1.436&1.628&2.031&1.736 \\
&$\pi$1h&5.448&5.458&5.976&5.560&4.653&5.661&6.576&5.749 \\
\etab\erule
\end{table}

\FloatBarrier

\end{document}